\documentclass[fleqn,usenatbib]{mnras}

\usepackage{newtxtext,newtxmath}

\usepackage[T1]{fontenc}
\usepackage{ae,aecompl}

\usepackage{graphicx}	
\usepackage{amsmath}	

\newcommand{\kms}{{\rm\,km\,s^{-1}}}
\newcommand{\kmskpc}{{\rm\,km\,s^{-1}{kpc}^{-1}}}
\newcommand{\grad}{{^\circ}}

\title[Orbital trapping by bar resonances]{Effect of orbital trapping
by bar resonances in the local U-V velocity field}

\author[E. Moreno et al.]{
Edmundo Moreno,$^{1}$\thanks{E-mail: edmundo@astro.unam.mx}
Jos\'e G. Fern\'andez-Trincado,$^{2,3}$
William J. Schuster,$^{4}$
Angeles P\'erez-Villegas,$^{4}$ \thanks{E-mail: mperez@astro.unam.mx}
\newauthor
\,Leonardo Chaves-Velasquez$^{5,6,7}$
\\
$^{1}$Instituto de Astronom\'ia, Universidad Nacional Aut\'onoma de
M\'exico, Apdo. Postal 70-264, Ciudad Universitaria CDMX 04510,
M\'exico\\
$^{2}$Instituto de Astronom\'ia y Ciencias Planetarias, Universidad de Atacama, Copayapu 485, Copiap\'o, Chile\\
$^{3}$Instituto de Astronom\'ia, Universidad Cat\'olica del Norte, Av. Angamos 0610, Antofagasta, Chile\\
$^{4}$Instituto de Astronom\'ia, Universidad Nacional Aut\'onoma de
M\'exico, Apdo. Postal 106, 22800 Ensenada, B.C., M\'exico\\
$^{5}$University of Nari\~no Observatory, Universidad de Nari\~no, Sede VIIS, Avenida Panamericana, Pasto, Nari\~no, Colombia\\
$^{6}$Departamento de F\'isica de la Universidad de Nari\~no, Torobajo Calle 18 Carrera 50, Pasto, Nari\~no, Colombia\\
$^{7}$Instituto de Radioastronom\'ia y Astrof\'isica, Universidad Nacional Aut\'onoma de M\'exico, Apdo. postal 3-72, Morelia Mich. 58089, M\'exico
}

\date{Accepted XXX. Received YYY; in original form ZZZ}

\pubyear{2021}

\begin{document}
\label{firstpage}
\pagerange{\pageref{firstpage}--\pageref{lastpage}}
\maketitle

\begin{abstract}

The effects in the local U-V velocity field due to orbital
trapping by bar resonances have been studied computing fifteen
resonant families in a non-axisymmetric Galactic potential, considering
the bar's angular velocity between 35 and 57.5 $\kmskpc$. Only cases
in the low, 37.5, 40 $\kmskpc$, and high, 55, 57.5 $\kmskpc$,
velocity ranges give trapping structures that have some similarity with
observed features in the velocity distribution. 
The resulting structures in the local U-V plane form resonant bands
appearing at various levels in velocity V. Cases with angular velocity
40 and 55 $\kmskpc$ show the greatest similarity with observed branches.
Our best approximation to the local velocity field by orbital
trapping is obtained with a bar angular velocity of 40 $\kmskpc$
and a bar angle of 40$\grad$. With this solution, three main observed 
features can be approximated: i) the Hercules branch at V=$-50$ $\kms$
produced by the resonance 8/1 outside corotation, and the close
features produced by resonances 5/1 and 6/1, ii) the newly detected
low-density arch at V $\simeq$ 40 $\kms$ produced approximately by the
resonance 4/3, iii) the inclined structure below the Hercules branch,
also observed in the $\textit{Gaia}$ DR2 data, produced by tube orbits
around Lagrange point $L_5$ at corotation. 
Some predicted contributions due to orbital trapping in regions of the
U-V plane corresponding to the Galactic halo are given, which could
help to further restrict the value of the angular velocity of the
Galactic bar. No support by orbital trapping is found for the Arcturus
stream at V $\approx$ $-100$ $\kms$.

\end{abstract}




\section{Introduction}

There is an extensive study of observed structures in the U-V
velocity field in the solar neighbourhood, beginning with data 
obtained by the satellite $\textit{Hipparcos}$   
\citep{1998AJ....115.2384D,1999A&A...341..427A,1999A&AS..135....5C},
combined with other data sources
\citep{2005A&A...430..165F,2008A&A...490..135A}, and employing
data from $\textit{Gaia}$ DR2 \citep{2018A&A...616A..11G}.
 Four main structures,
or branches, Sirius, Coma Berenices, Hyades-Pleiades, and Hercules,
emerged from these analyses \citep{1999MNRAS.308..731S,
2008A&A...490..135A}, with several internal concentrations, or
moving groups. Possible mechanisms responsible for the existence of
these structures have been suggested: 
cluster remnants, merger events (probably the case of the
Arcturus stream \citep{2019A&A...631A..47K}), dynamical effects of the bar and spiral arms,
and interaction with bar and/or spiral arms resonances.

 There is a large scatter in
age and metallicity of stars forming these structures
\citep{1998AJ....115.2384D,1999A&AS..135....5C,2005A&A...430..165F, 
2007ApJ...655L..89B,2008A&A...490..135A}, which shows that they are
not entirely remnants of particular dispersed star clusters nor
share the same formation history. Thus, the
studies have focused on a dynamical origin: merger or resonance
effects. First evidences of orbital trapping by bar resonances as a
more appropriate mechanism were given by 
\citet{1998AJ....115.2384D,1999ApJ...524L..35D,2000AJ....119..800D}
and \citet{2001A&A...373..511F}, and there are several studies 
following in this direction \citep{2010MNRAS.405..545G,
2016AstL...42..228B,2017ApJ...840L...2P,2018MNRAS.474...95H,
2018MNRAS.477.3945H,2019A&A...626A..41M}.
The effects of spiral arms alone or combined with the bar 
have also been considered \citep{2005AJ....130..576Q,
2007A&A...467..145C,2009ApJ...700L..78A,2010HiA....15..192A,
2011MNRAS.418.1423A,2018A&A...615A..10M,2018ApJ...863L..37M,
2019MNRAS.490.1026H,2020ApJ...888...75B}.
 Effects of transient spiral
arms \citep{2004MNRAS.350..627D} and transient features created by a
bar \citep{2010MNRAS.407.2122M} are other possible mechanisms to
explain local features. Other models consider perturbations due to a
merger event \citep{2009MNRAS.396L..56M,2019A&A...631A..47K}.

Some features in the local U-V field are reproduced by all these
proposed models, but further analyses are needed. Here we consider
in some detail the possible effects of orbital trapping due
to resonances created by a bar on the Galactic plane, with emphasis in
the U-V region occupied by the four main branches. Fifteen main 
resonant orbital families are analysed in a wide interval of bar
angular velocities, and their contributions to the local U-V velocity
field are computed considering their stable orbital sections.
The non-axisymmetric Galactic model employed is described
in Section~\ref{modelo}, with the adopted properties of the Galactic
bar. The fifteen resonant families are considered in 
Section~\ref{reson}, and their resulting contributions to the U-V
field are given in Section~\ref{uvperiodicas} and \ref{uvatrap}, along
with a comparison with known results. Our conclusions appear in
Section~\ref{concl}.

\section{The Non-axisymmetric Galactic Model}
\label{modelo}

The Galactic model used has three axisymmetric components and a
Galactic bar. It is based on the Galactic axisymmetric model of
\citet{1991RMxAA..22..255A}, which consists of a spherical bulge and
a disk, both of the Miyamoto-Nagai type \citep{1975PASJ...27..533M}, 
and a spherical dark halo with mass inside distance $r$ from the
Galactic centre given by

\begin{equation}
  M_h(r)=\frac{M_3(r/a_3)^{s_1}}{1+(r/a_3)^{s_2}},
\label{mhalo}
\end{equation}

with $s_1$=2.02, $s_2$=1.02 and $M_3$, $a_3$ and other parameters of
the bulge and disk listed in table 1 of \citet{1991RMxAA..22..255A}.

This axisymmetric model was rescaled to the Sun's galactocentric
distance $R_0$=8.3 kpc and the Local Standard of Rest (LSR) velocity 
$\Theta_0$=239 $\kms$, as determined in \citet{2011AN....332..461B}.
After this transformation, from the resulting total mass of 
1.62$\times 10^{10} M_{\odot}$ in the bulge component, a
1$\times 10^{10} M_{\odot}$ Galactic prolate bar was built, with the
rest of its mass remaining as a diminished spherical bulge.
The assumed mass of the bar is around values proposed by
\citet{1996MNRAS.283..149Z}, \citet{1999ApJ...524..112W}, 
\citet{2010MNRAS.405..545G}. This prolate bar approximates Model S of
\citet{1998ApJ...492..495F} of \textit{COBE}/DIRBE observations of
the Galactic centre; its mass stratification is similar, i.e.
equidensity prolate surfaces with the same eccentricity. It has a
density law of the form

\begin{equation}
 \rho(R_s)=
 \begin{cases}
 \rho_0 sech^2(R_s), & \text{$R_s\leq R_B$}\\
 \rho_0 sech^2(R_s)e^{-(R_s-R_B)^2/h_B^2}, & \text{$R_s\geq R_B$}
 \end{cases}
\label{denprol}
\end{equation}

\begin{equation}
 R_s=\left\{\frac{{x^\prime}^2}{a_{{x_p^\prime}}^2} +
 \frac{{y^\prime}^2+{z^\prime}^2}{a_{{y_p^\prime}}^2}\right\}^{1/2}
\label{Rs}
\end{equation}

\noindent with $\rho_0$ depending on the mass and size of the bar;
$x^\prime,y^\prime,z^\prime$ are Cartesian coordinates along the axes of the bar with the
$x^\prime$-axis on the major axis, the axes $x^\prime$
and $y^\prime$ lie on the Galactic plane; $a_{{x_p^\prime}}$=$a_{x^\prime}$,
$a_{{y_p^\prime}}$=$\frac {1}{2}(a_{y^\prime}+a_{z^\prime})$ with
scale lengths ($a_{x^\prime}$,$a_{y^\prime}$,$a_{z^\prime}$)=(1.66,\,0.62,\,0.43)kpc 
and an effective major semiaxis of the bar $a_B$=3.06 kpc
corresponding to $R_0$=8.3 kpc; 
$R_B$=$a_B/a_{x^\prime}$, and $h_B$=$h/a_{x^\prime}$ with
$h$=0.46 kpc. The eccentricity of the bar is
$e$=$(1-a_{{y_p^\prime}}^2/a_{{x_p^\prime}}^2)^{1/2}$. If $a_B$ is changed, as considered
in Section~\ref{vang} below, the scale lengths are changed in the same
proportion, and the eccentricity remains the same. The gravitational
potential of the bar is given in \citet{2004ApJ...609..144P}.

\subsection{Angular velocity and Size of the bar}
\label{vang}

The estimated values of the Galactic bar's angular velocity,
$\Omega_{\rm b}$, cover a wide interval: $\approx$ 25 -- 65 $\kmskpc$
\citep[and references therein]{2011MSAIS..18..185G,2016ARA&A..54..529B}.
From gas dynamics in the inner Galactic region, large angular velocities
above 50 $\kmskpc$ have been obtained \citep{1999MNRAS.304..512E,
1999A&A...345..787F,2003MNRAS.340..949B}. The corotation radius,
$R_{\rm cr}$, in 
these cases lies in the range $\approx$ 3 -- 4.5 kpc.
Other hydrodynamic simulations have estimated lower values 
$\Omega_{\rm b}$ $\approx$ 30 -- 42 $\kmskpc$ 
\citep{1999ApJ...524..112W,2008A&A...489..115R,
2015MNRAS.454.1818S,2016ApJ...824...13L}, with $R_{\rm cr}$ in the
range $\approx$ 5 -- 7.5 kpc. Dynamical models relating the U,V
velocity field in the solar neighbourhood with resonant orbits near
the outer Lindblad resonance, also give a wide scatter in
$\Omega_{\rm b}$. Modelling a bimodality in this local velocity field,
\citet{1999ApJ...524L..35D,2000AJ....119..800D} obtains
53 $\pm$ 3 $\kmskpc$. From the kinematics of the Hercules stream,
\citet{2014A&A...563A..60A} find a relation for $\Omega_{\rm b}$
which gives $\Omega_{\rm b}$ $\simeq$ 55 $\kmskpc$, scaled to our
adopted ($R_0$,$\Theta_0$)=(8.3 kpc,\,239$\kms$).
\citet{2017ApJ...840L...2P} give an interpretation of the Hercules
stream as orbital interactions with the corotation resonance,
resulting in $\Omega_{\rm b}$=39 $\kmskpc$. With 2:1 resonant orbits,
\citet{2016AstL...42..228B} fit four features in the Hercules and
Wolf 630 streams and find 45 -- 55 $\kmskpc$. Also, analysing the
position of the Hercules stream in velocity space as a function of
radius in the outer Galaxy, \citet{2017MNRAS.466L.113M} find that
$\Omega_{\rm b}$ $\geq$ 1.8$\Omega_0$, with
$\Omega_0$=$\Theta_0$/$R_0$, thus favoring fast bar models. In a
previous study \citet{2017MNRAS.465.1443M} give the same conclusion
exploring the response of stars in the solar neighbourhood to slow and
fast bars, showing that slow bar models, as those supported by
hydrodynamic simulations, are unable to reproduce the bimodality
observed in the local U-V velocity field; thus they conclude that
in order to explain this bimodality with a slow bar, an alternative
explanation should be found.
In this respect, \citet{2018MNRAS.477.3945H} show that a slow bar
can reproduce a Hercules-like feature if the bar potential includes
an m=4 Fourier component. Dynamical and kinematic models also predict
low and large values of $\Omega_{\rm b}$. \citet{2017MNRAS.465.1621P}
obtain $\Omega_{\rm b}$=39$\pm$3.5$\kmskpc$,
\citet{2007ApJ...664L..31M} $\Omega_{\rm b}$ $\geq$ 1.8$\Omega_0$,
\citet{2019MNRAS.488.4552S} $\Omega_{\rm b}$=41$\pm$3$\kmskpc$.
A slow bar with $\Omega_{\rm b}$ around 40 $\kmskpc$ is supported by
recent studies \citep{2019A&A...626A..41M,
2019A&A...632A.107M,2019MNRAS.488.4552S,2019MNRAS.489.3519C,
2019MNRAS.490.4740B}. 
 
\begin{figure}
\includegraphics[width=\columnwidth]{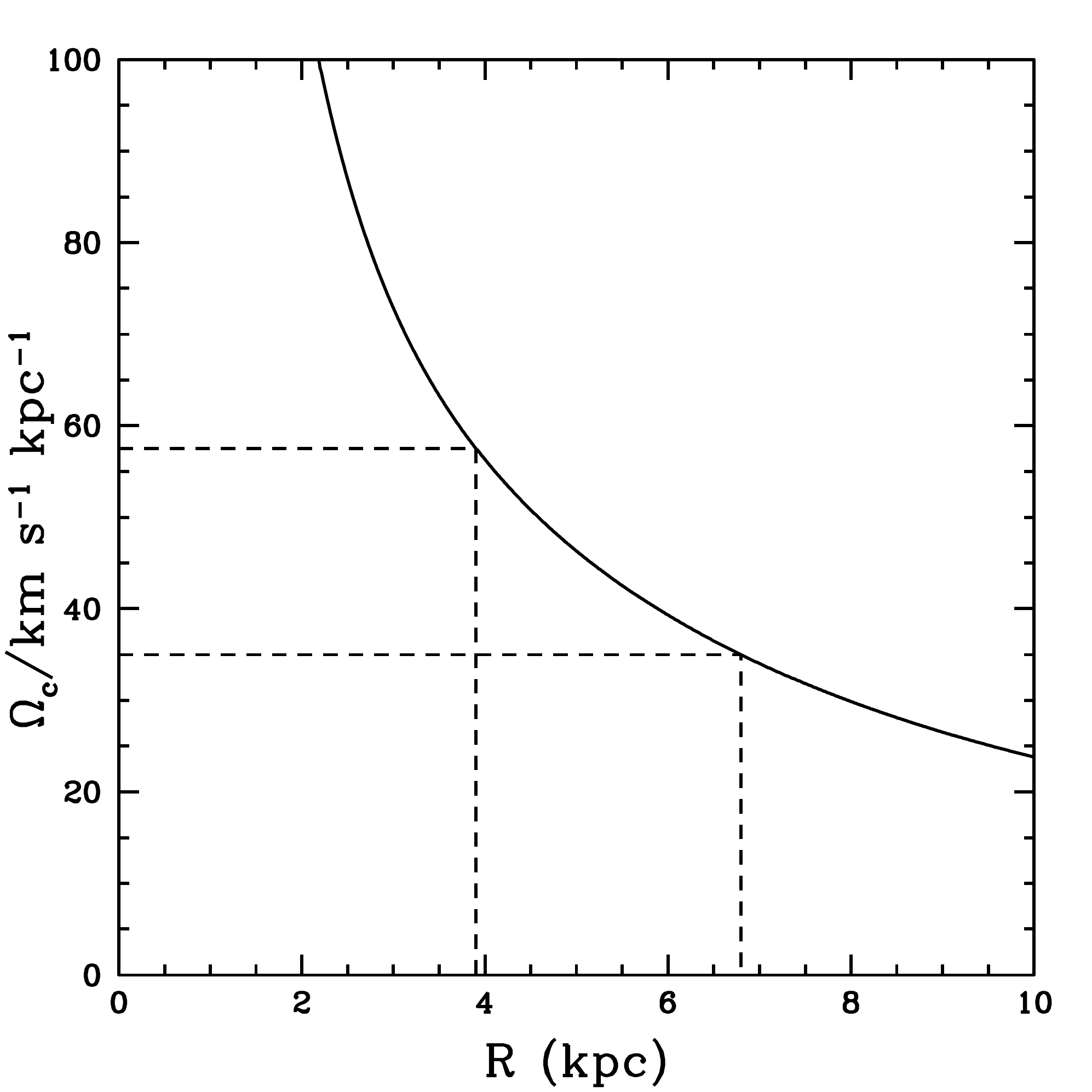}
\caption{Angular velocity of circular orbits in the scaled
axisymmetric Galactic model. The corotation radii 
$R_{\rm cr}$=6.8,\,3.9 kpc are shown at 
$\Omega_{\rm b}$=35,\,57.5 $\kmskpc$, respectively.
 The bar's major semiaxis $a_B$ is
obtained with the resulting corotation radius $R_{\rm cr}$ assuming
$a_B$=$R_{\rm cr}$/1.2}
\label{figura1}
\end{figure}

To represent these results of long and slow, and short and fast bars,
in our computations the bar's major semiaxis is obtained from
$\Omega_{\rm b}$ assuming the relation between the corotation radius
and major semiaxis $\mathcal{R}$=$R_{\rm cr}/a_B$=1.2 \citep{BT08}.
Ten values of the angular velocity of the bar are considered, between
35 and 57.5 $\kmskpc$ in steps of 2.5 $\kmskpc$. Focusing on this
interval, Fig.~\ref{figura1} shows
the angular velocity of circular orbits, $\Omega_{\rm c}$, in the 
scaled axisymmetric Galactic model as a function of distance $R$ on the
Galactic plane. With $\Omega_{\rm b}$=35,\,57.5 $\kmskpc$, the
corresponding values of the corotation radius are $R_{\rm cr}$=
6.8,\,3.9 kpc. 

\subsection{Orientation of the bar's major axis}
\label{ang}

As in the case of the bar's angular velocity, the determined present
value of the angle between the major axis of the bar and the
Sun-Galactic centre line, $\phi_B$, has a large uncertainty.
Some obtained values are:
20$\grad$ -- 25$\grad$ \citep{1999MNRAS.304..512E,2003MNRAS.340..949B},
25$\pm$4$\grad$ \citep{1999A&A...345..787F}, 
34$\grad$ \citep{1999ApJ...524..112W}, 
10$\grad$ -- 70$\grad$ \citep{2000AJ....119..800D},
43$\pm$7$\grad$ \citep{2000MNRAS.317L..45H}, 
44$\pm$10$\grad$ \citep{2005ApJ...630L.149B}, 
22$\pm$5$\grad$ \citep{2005MNRAS.358.1309B}
20$\grad$ -- 45$\grad$ \citep{2007ApJ...664L..31M}, 
24$\grad$ -- 27$\grad$ \citep{2007MNRAS.378.1064R}
20$\grad$ -- 35$\grad$ \citep{2008A&A...489..115R}, 
15$\pm$13$\grad$ \citep[other $\phi$ values are given in their table 1]{2009A&A...498...95V},
13$\grad$ \citep{2012A&A...538A.106R},
27$\grad$,28$\grad$ \citep{2013MNRAS.435.1874W,2017MNRAS.465.1621P}.

In our analysis we take ten values of $\phi_B$ in the range
5$\grad$ -- 50$\grad$, in steps of 5$\grad$,
this with the purpose to take into account the effect of the bar angle
in the velocity distribution of the solar neighbourhood.

\section{Bar resonances on the Galactic plane}
\label{reson}

\subsection{Resonant Orbital Families and their Stability}
\label{fam}

With the non-axisymmetric Galactic model described in
 Section~\ref{modelo},
for each value of $\Omega_{\rm b}$ we computed several resonant orbital
families on the Galactic plane,
generated by the bar component. These families consist of
periodic orbits which can be represented in a diagram plotting a
characteristic orbital energy versus the orbital Jacobi constant
 $E_{\rm J}$.
This diagram was employed in \citet{2015MNRAS.451..705M}, with the
characteristic orbital energy defined by ($E_{\rm min}+E_{\rm max}$)/2,
with $E_{\rm min}$, $E_{\rm max}$ the minimum and maximum energies
per unit mass along each orbit, computed with respect to the Galactic
\textit{inertial} reference frame. $E_{\rm J}$ has a constant value in 
the \textit{non-inertial} reference frame where the bar is at rest,
and is given by

\begin{equation} E_J = \frac{1}{2}{\bf v^\prime}^2 + \Phi_0({\bf r^\prime}) +
\Phi_{\rm b}({\bf r^\prime}) - \frac{1}{2}{\Omega^2_{\rm b}}({x^\prime}^2 + {y^\prime}^2),
\label{J}
\end{equation}

\noindent with $\bf r^\prime$=($x^\prime,y^\prime,z^\prime$)
and $\bf v^\prime$ position and velocity in this frame; $\Phi_0$, 
$\Phi_{\rm b}$ are the potentials due to the axisymmetric mass
distribution and the bar, respectively. The velocity $\bf v$,
and energy per unit mass $E$ with respect to the Galactic inertial
frame are

\begin{equation} {\bf v} = {\bf v^\prime} + {\bf \Omega_b} \times
{\bf r^\prime}, 
\label{Vel}
\end{equation}

\begin{equation} E = \frac{1}{2}{\bf v}^2 + \Phi_0({\bf r^\prime}) +
\Phi_{\rm b}({\bf r^\prime}),
\label{E}
\end{equation}

\noindent with the vector $\bf \Omega_b$ pointing to the
South Galactic pole.

By computing several Poincar\'e section diagrams on the Galactic plane, a
starting periodic orbit was located in each family and from this orbit
other periodic orbits associated to a family were successively found,
with a typical separation $\Delta E_{\rm J}$=10$^2$ km$^2$ s$^{-2}$,
using a Newton--Raphson method \citep{PTVF92}.

\citet{2015MNRAS.451..705M} analysed eleven resonant families,
numbered I to XI in their figure 6, in a Galactic potential similar to
the one employed here, taking a fixed value
$\Omega_{\rm b}$=55 $\kmskpc$.
In the present study we consider their families I to IX plus six other
families numbered 1 to 6, giving a total number of fifteen families
computed in each of the ten values considered for $\Omega_{\rm b}$,
except family 1 which was employed only with
 $\Omega_{\rm b}$=35,\,37.5,\,40 $\kmskpc$,
not contributing with other values of the bar's
angular velocity in the analysis made in Sections~\ref{uvperiodicas},
\ref{uvatrap}.

\begin{figure}
\includegraphics[width=\columnwidth]{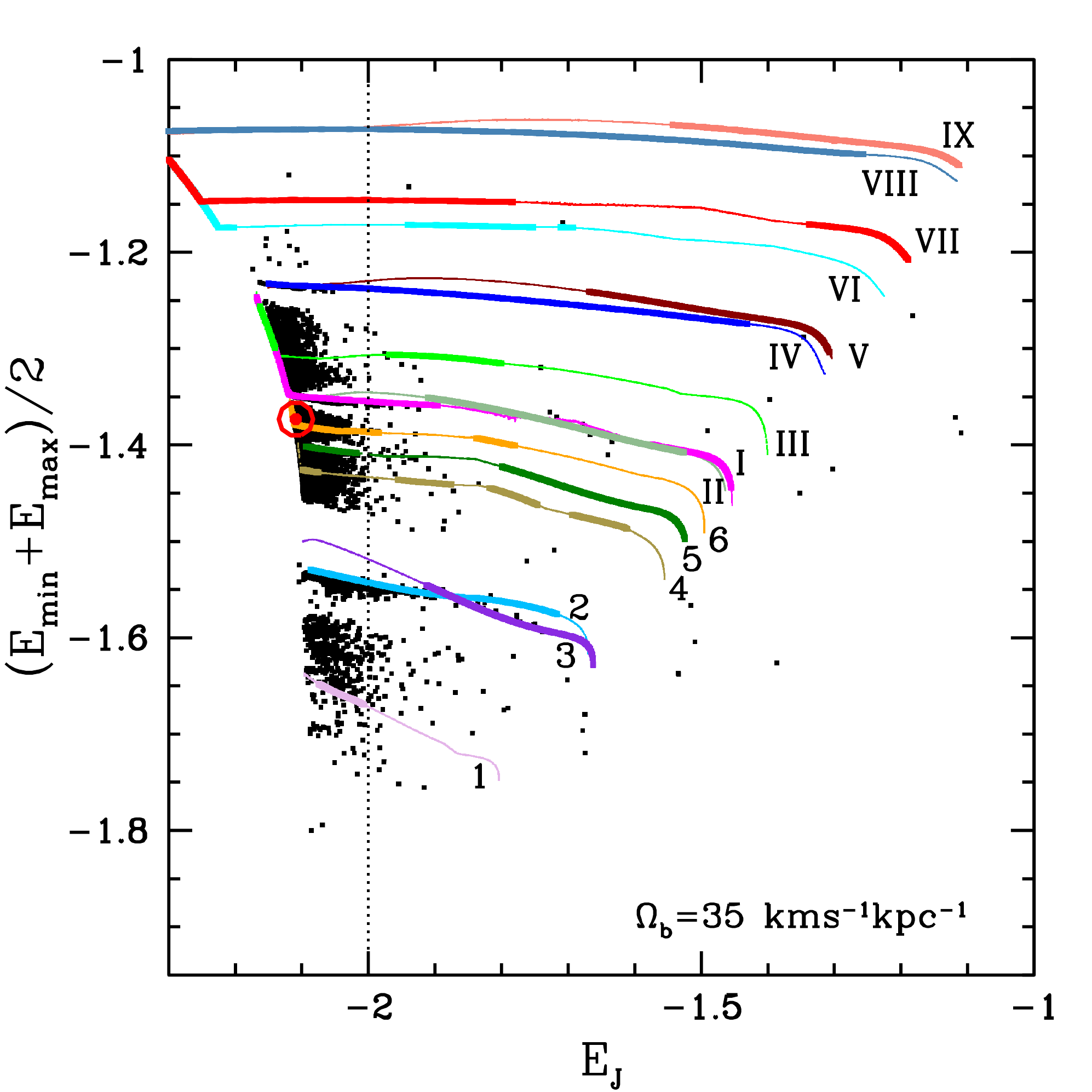}
\caption{Characteristic energy versus $E_{\rm J}$, in units of 
10$^5$ km$^2$ s$^{-2}$ in both axes, showing the fifteen resonant 
orbital families on the Galactic plane considered in our analysis.
The angular velocity of the bar is 35 $\kmskpc$ in this figure.
The black points correspond to orbits of 2 $\times$ 10$^4$ stars taken
 at random
within a 2 kpc-radius solar neighbourhood in $\textit{Gaia}$ DR2.
The thick and thin parts of a curve are respectively the stable and
unstable orbital sections of the family.
The big red circle with a central dot shows the position of the Sun.}
\label{figura2}
\end{figure}

\begin{figure}
\includegraphics[width=\columnwidth]{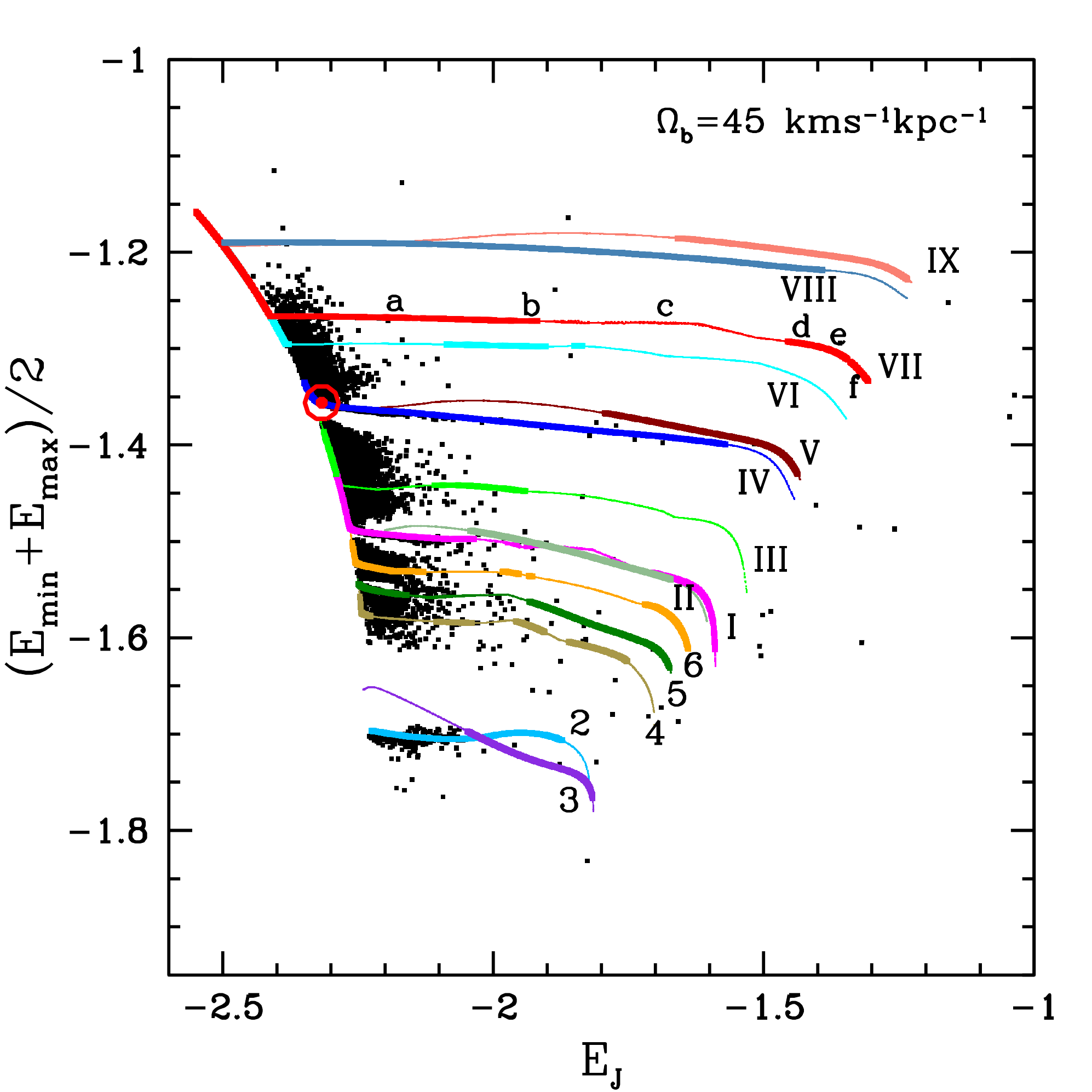}
\caption{As in Fig.~\ref{figura2}, here with an angular velocity of
45 $\kmskpc$.}
\label{figura3}
\end{figure}

\begin{figure}
\includegraphics[width=\columnwidth]{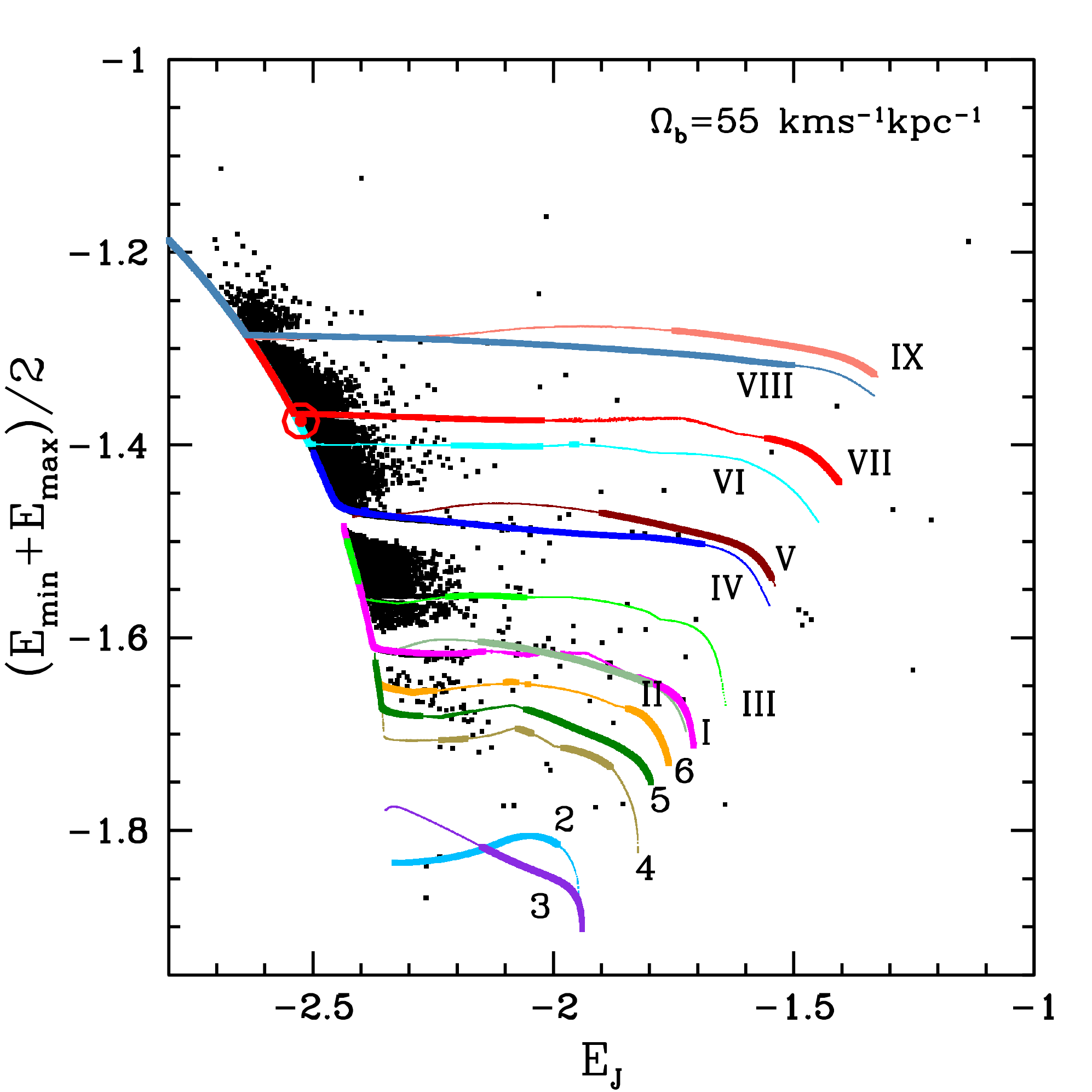}
\caption{As in Fig.~\ref{figura2}, here with an angular velocity of
55 $\kmskpc$. Note the downwards shift of the resonant families with
respect to their positions in Fig.~\ref{figura2}.} 
\label{figura4}
\end{figure}

Figs.~\ref{figura2},~\ref{figura3},~\ref{figura4} show all the
families in diagrams of characteristic energy versus $E_{\rm J}$, in
particular for $\Omega_{\rm b}$=35,\,45,\,55 $\kmskpc$. Each family
has its corresponding number. In our analysis we consider
only the prograde regions of each family, i.e. rotating with respect
to the inertial Galactic frame in the same sense as the bar. 
The black points shown in these figures correspond to orbits of
2 $\times$ 10$^4$ stars taken at random from stars within a
2 kpc-radius solar neighbourhood listed in
$\textit{Gaia}$ DR2 \citep{2018A&A...616A...1G}.
 The majority of these sampled
stars are prograde and distribute close to the left boundary in a
diagram. Note in these diagrams how resonant families shift downwards,
i.e. their member orbits decrease their distance from the Galactic
centre, as $\Omega_{\rm b}$ is increased, and there is a change in
the distribution of the sampled stars among these families;
the major change occurs towards the inner Galactic region. 
In Figs.~\ref{figura2},~\ref{figura3},~\ref{figura4}
the big red circle with a central dot shows the position of the Sun,
which approximately maintains its level in characteristic energy.
Thus, due to the shift of resonances, depending on $\Omega_{\rm b}$
there are different resonant families around this energy level which
may be important for orbital trapping in the solar neighbourhood.

 In general the sampled stars have
three-dimensional orbits, whereas the resonant families represent
two-dimensional orbits on the Galactic plane. As shown in
\citet{2015MNRAS.451..705M}, the motion parallel to the Galactic
plane of three-dimensional orbits can be trapped by two-dimensional
resonant families on this plane, even if these orbits reach
high altitudes. This orbital trapping is a central point in our
analysis presented in the following sections.

Orbital trapping can take place
where the member periodic orbits of a given family are stable.
The orbital stability of all the resonant families was analysed
following the treatment given by \citet{1965AnAp...28..992H} and
\citet{C02}, for all the values of the bar's angular velocity.
In particular for the cases $\Omega_{\rm b}$= 35, 45, 55 $\kmskpc$
presented in Figs.~\ref{figura2},~\ref{figura3},~\ref{figura4}, the
results of the stability analysis are shown with the thickness
of the corresponding curve of a family: the thick and thin parts
of a curve are respectively the stable and unstable orbital sections
of the family.
There are some agglomerations of black points around stable parts
of several families; particularly strong, for example, in families
 2, 5, 6, I, IV.
In all these cases, orbital trapping appears to be important.

\begin{figure}
\includegraphics[width=\columnwidth]{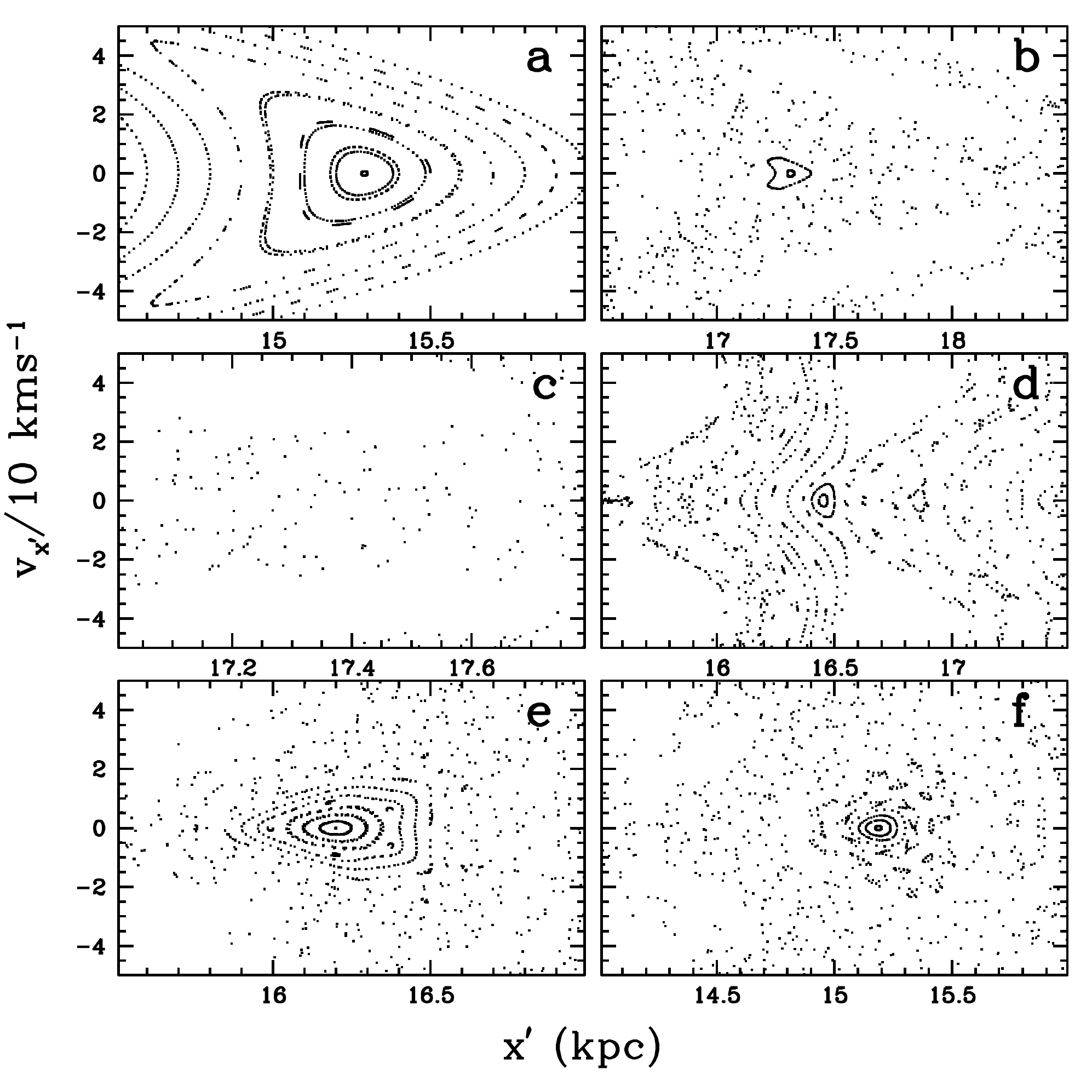}
\caption{Poincar\'e diagrams at the marked points a -- f on the
curve of family VII in Fig.~\ref{figura3}. These diagrams show a
region around the perpendicular crossing of the member periodic orbit
in the family with the bar's long axis. A wide region of islands
appears in the marked stable points a,e. These islands do not appear
in the unstable point c. In the marked points b,d,f near a
stable-unstable transition, the island region is small.}
\label{figura5}
\end{figure}

To illustrate the stable-unstable behavior along a family curve,
in particular Fig.~\ref{figura5} shows Poincar\'e diagrams 
$V_{x^{\prime}}$ vs $x^{\prime}$, with $V_{x^{\prime}}$
velocity in the non-inertial reference frame of the bar, for family
 VII in Fig.~\ref{figura3} at the marked points a -- f on
its curve. These diagrams show only a region around the intersection
point where the corresponding member periodic orbit in the family
crosses perpendicularly the bar's long axis, i.e. the
$x^{\prime}$ axis. In the marked stable points a,e a wide region of
islands is obtained around the central periodic orbit in the family.
In the marked points b,d,f the region of islands has almost disappeared
because these points are near the transition to orbital instability.
The marked point c lies in this region of instability and there are no
islands around the periodic orbit.
Values of the velocity $V_{x^{\prime}}$ on the island regions similar
to the ones shown in this Fig.~\ref{figura5}, will be employed in
Section~\ref{uvatrap}.

\subsection{ Periodic Orbits}
\label{orbs}

\begin{figure}
\includegraphics[width=\columnwidth]{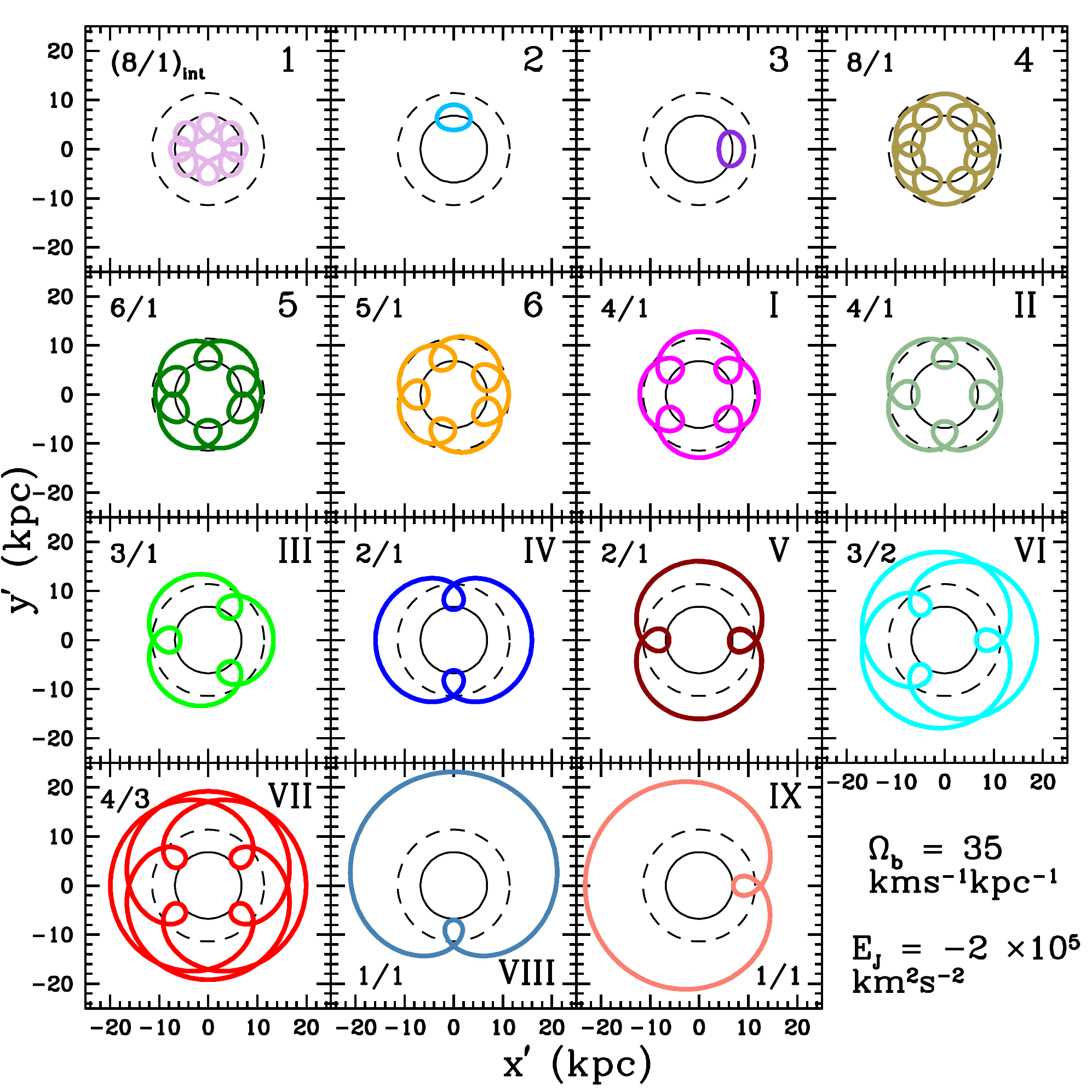}
\caption{The periodic orbits in each resonant family in 
Fig.~\ref{figura2} at $E_{\rm J}$ = $-2$, i.e. 
$E_{\rm J}$ = $-2$ $\times$ 10$^5$ km$^2$ s$^{-2}$ (dotted vertical
line in that figure). The colour of an orbit
is the same of the family. The $x^{\prime}$ and $y^{\prime}$ axes
lie on the Galactic plane, along the major and minor axes of the bar,
respectively. The black continuous circle shows the position of the
corotation resonance, and the discontinuous circle the outer Lindblad
resonance. The rotation number $n/m$ (see main text) of the resonant
family is given in the corresponding panel.}
\label{figura6}
\end{figure}

By taking in Fig.~\ref{figura2} a vertical line (dotted line) at
$E_{\rm J}$ = $-2$ $\times$ 10$^5$ km$^2$ s$^{-2}$, the corresponding periodic
orbits at the intersections with the resonant families are plotted
in Fig.~\ref{figura6}. The orbits have the same colour as their parent
families. As stated above, the $x^{\prime}$ and $y^{\prime}$ axes on
 the Galactic
plane point along the major and minor axes of the bar, respectively,
and corotate with the bar. The Galactic centre is at the origin.
The continuous black circle shows the position of the corotation
resonance, and the discontinuous circle the outer Lindblad resonance.
\citet{2001A&A...373..511F} shows some periodic orbits obtained with
a quadrupole potential bar.

In each panel of Fig.~\ref{figura6} we give the ratio $n/m$,
or rotation number \citep{C02}, of the periodic orbit in the
non-inertial frame; $n$ is the number of oscillations in the radial
direction when the orbits makes $m$ oscillations around the Galactic
centre. The rotation number tends to infinity as we approach
corotation resonance. For family 1 this rotation number is denoted with a subindex
\textit{int}, meaning interior to corotation resonance. The given
rotation number in a family will change when its orbits change from
prograde to retrograde, i.e. when they cross the Galactic centre and
the curves in Figs.~\ref{figura2},~\ref{figura3},~\ref{figura4} are
extended to the right sides.
Families II,V,IX emerge as unstable bifurcations of stable sections
of families I,IV,VIII, respectively, and have similar rotation
numbers 4/1, 2/1, 1/1. In the following we refer to a family by its
number or stating the corresponding $n/m$ resonance.

\begin{figure}
\includegraphics[width=\columnwidth]{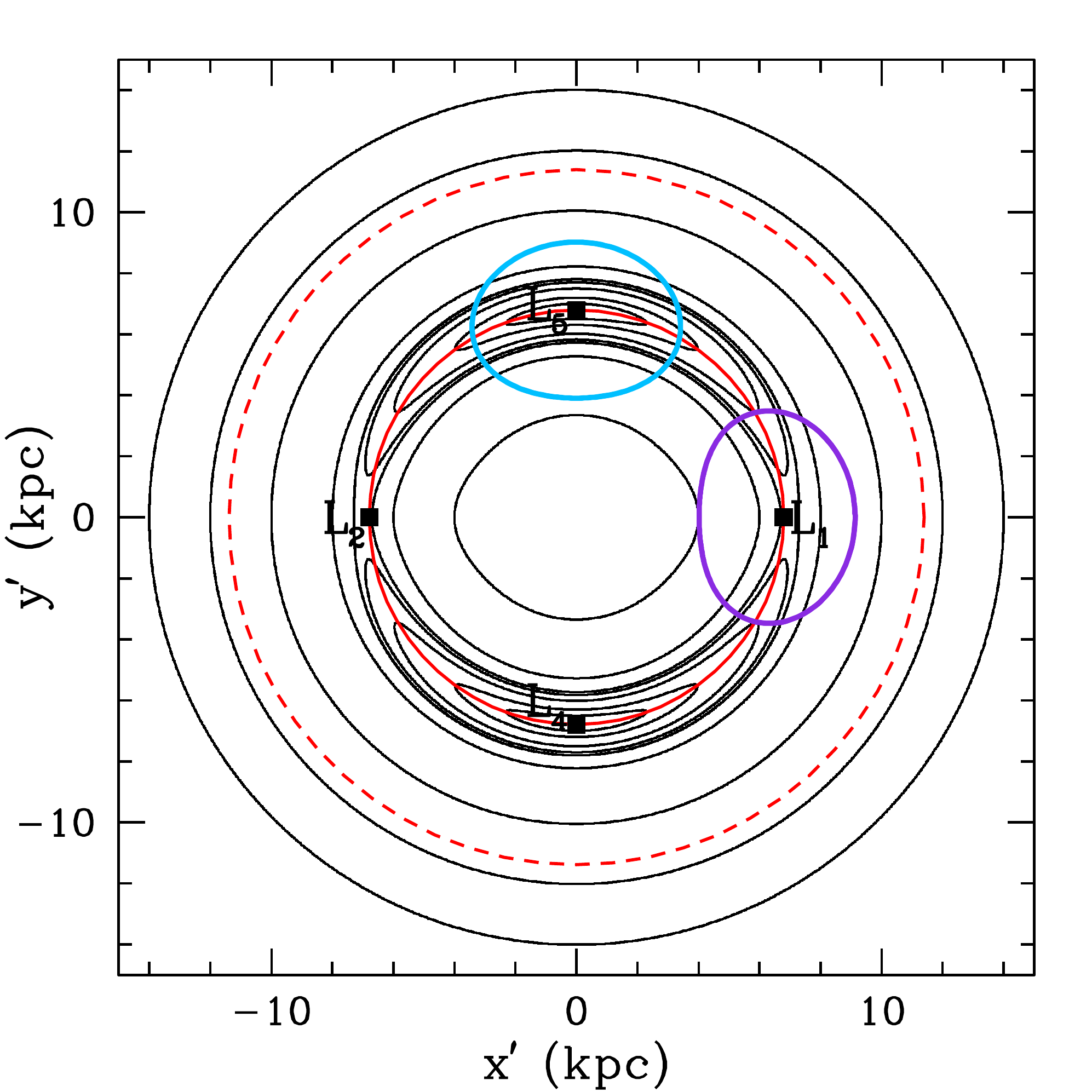}
\caption{The periodic orbits of families 2 and 3 (blue and purple
curves, respectively) in Fig.~\ref{figura6}
along with some zero-velocity curves (in black) computed with 
$\Omega_{\rm b}$= 35 $\kmskpc$. These orbits encircle the
Lagrange points $L_5$, $L_1$. The corotation resonance is shown by
the red continuous circle; the outer Lindblad resonance lies at the
discontinuous red circle. Lagrange points $L_2$, $L_4$ are also shown.} 
\label{figura7}
\end{figure}

In particular, the periodic orbits of families 2 and 3 in
Fig.~\ref{figura6} are plotted in Fig.~\ref{figura7} showing some
zero-velocity curves computed in the Galactic model with
$\Omega_{\rm b}$= 35 $\kmskpc$. With the notation given in figure 3.14
of \citet{BT08}, these periodic orbits in families
2,3 lie around Lagrange points $L_5$, $L_1$, respectively (black dots
in Fig.~\ref{figura7}); the
symmetric orbits around Lagrange points $L_4$, $L_2$ also belong to
these families. In Fig.~\ref{figura7} the red continuous and
discontinuous circles, show the corotation resonance and outer
Lindblad resonance, respectively. 

\section{The local U-V velocity field of resonant families crossing
the solar neighbourhood}
\label{uvperiodicas}

With the stable sections of each resonant family, the next step is to
compute the corresponding U-V velocity field of the \textit{periodic}
orbits in each section if they cross the solar neighbourhood.
This gives a first estimate of their possible contribution to the
observed local U-V field.
In Section~\ref{uvatrap} we will focus in the contributions of
\textit{trapped} orbits around stable periodic orbits.

A solar vicinity on the Galactic plane with radius 500 pc is
considered, at the Sun's galactocentric distance $R_0$=8.3 kpc and
in the range of values of the angle $\phi_B$, between the major axis
of the bar and the Sun-Galactic centre line, given in Section~\ref{ang}.
If a periodic orbit crosses the solar vicinity, it is analysed at close
division points, computing at each one the velocity components U,V
with respect to the Sun. This is done with the point's position 
($x^{\prime}$,$y^{\prime}$) and velocity 
($V_{x^{\prime}}$,$V_{y^{\prime}}$) with respect to
the non-inertial reference frame where the bar is at rest, the bar's
angular velocity $\Omega_{\rm b}$, the angle $\phi_B$, the LSR
velocity $\Theta_0$, and the peculiar Solar motion with respect to
the LSR, which was taken as 
$(U,V,W)_{\odot}$=(11.10,\,12.24,\,7.25)$\kms$
\citep{2010MNRAS.403.1829S}, with $U_{\odot}$ positive towards the
Galactic centre, as assumed for U with respect to the Sun.
Considering an inertial reference system with origin
at the Galactic centre, the $x$-axis pointing to the present
position of the Sun, and the $y$-axis pointing in the opposite
direction to Galactic rotation, the velocity components U,V of an
orbital point with respect to the Sun in terms of the velocity
($V_x$,$V_y$) in this system are 

\begin{equation}
U=-V_x+U_{\odot}=-(V_{x^{\prime}}+y^{\prime}\Omega_{\rm b})\cos\phi_B-
      (V_{y^{\prime}}-x^{\prime}\Omega_{\rm b})\sin\phi_B+U_{\odot} 
\label{u}
\end{equation}

\begin{equation}
V=-V_y-\Theta_0-V_{\odot}=
  (V_{x^{\prime}}+y^{\prime}\Omega_{\rm b})\sin\phi_B-
      (V_{y^{\prime}}-x^{\prime}\Omega_{\rm b})\cos\phi_B-\Theta_0-
      V_{\odot}
\label{v}
\end{equation}

As the Sun's vicinity is located in the first
quadrant of the ($x^{\prime}$,$y^{\prime}$) plane, in the resonant
families 6,III,VI,IX we take into account their orbital reflections
with respect to the $y^{\prime}$-axis, which give periodic orbits
belonging to the same families. Also, in family VIII its reflection
with respect to the $x^{\prime}$-axis belongs to this family.  
These reflected orbits exist due to the symmetries of the Galactic
potential.

\begin{figure}
\includegraphics[width=\columnwidth]{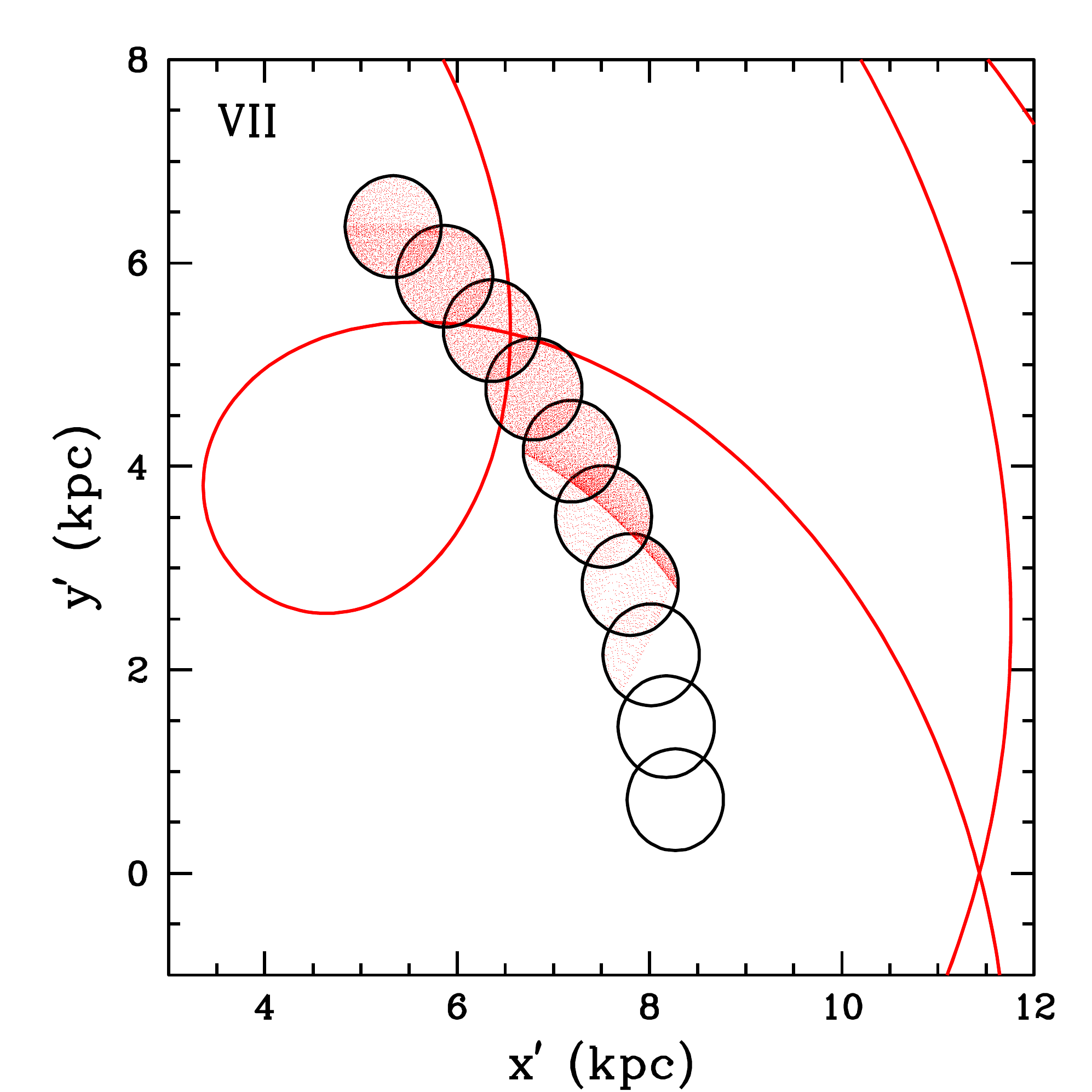}
\caption{Solar vicinities shown with circles with radius of 500 pc,
at different angles $\phi_B$, and inner orbital
points generated by a stable section of family VII, which is
symmetric with respect to the Galactic centre. A particular orbit
in the family is plotted. In this example,
$\Omega_{\rm b}$=50 $\kmskpc$.}
\label{figura8}
\end{figure}

\begin{figure}
\includegraphics[width=\columnwidth]{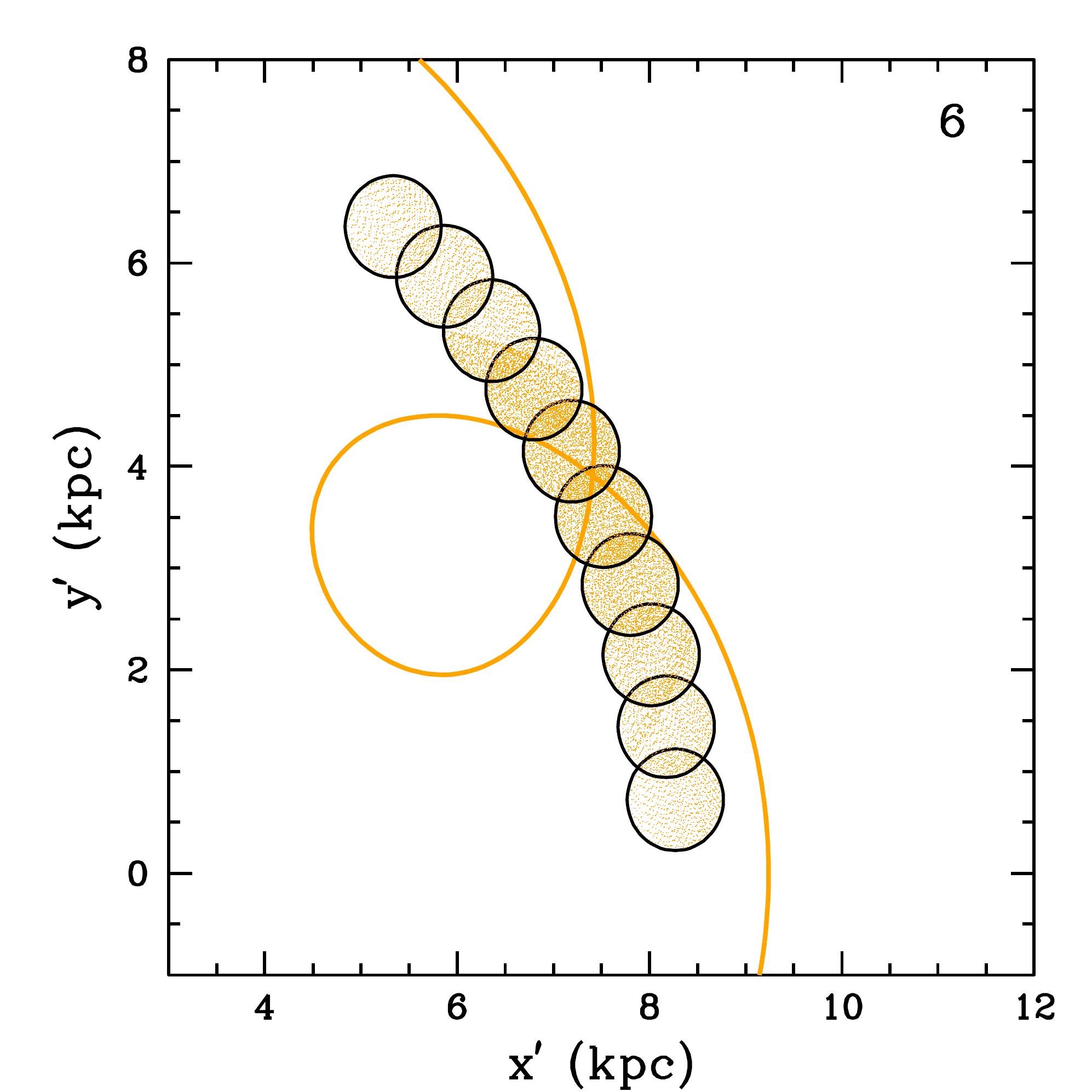}
\caption{As in Fig.~\ref{figura8}, here an example with a stable
section of family 6, with $\Omega_{\rm b}$=40 $\kmskpc$.}
\label{figura9}
\end{figure}

\begin{figure}
\includegraphics[width=\columnwidth]{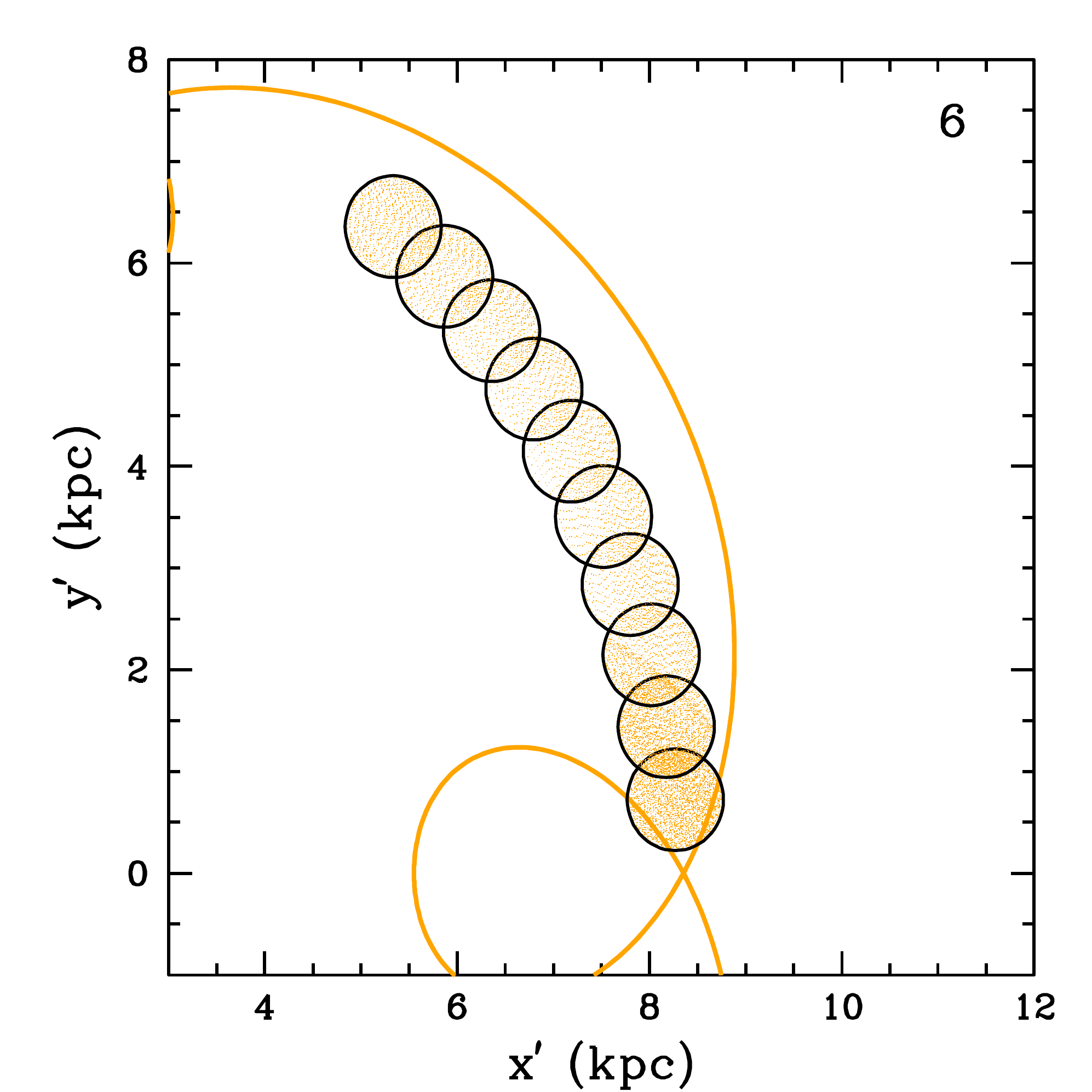}
\caption{Contributions of family 6 reflected with respect to the
$y^{\prime}$-axis, with $\Omega_{\rm b}$=40 $\kmskpc$.}
\label{figura10}
\end{figure}

An example of a resonant family which has symmetric orbits with respect
to the Galactic centre is given in Fig.~\ref{figura8}; it shows the
ten analysed solar vicinities and different orbital inner points due to
a stable section of family VII, with $\Omega_{\rm b}$=50 $\kmskpc$.
A particular orbit in the family is plotted. Depending on the
solar vicinity, more than one orbital crossing can contribute to the
inner points, thus increasing their number in some regions within a
given solar vicinity. If this vicinity is empty, there are no crossing 
points. Figs.~\ref{figura9},~\ref{figura10} show a similar
example now with a stable section of family 6, in the case
$\Omega_{\rm b}$=40 $\kmskpc$. Contributions of the orbital reflection
of this family with respect to the $y^{\prime}$-axis are shown in
Fig.~\ref{figura10}. Two particular orbits in the family are plotted
in these figures.

\begin{figure}
\includegraphics[width=\columnwidth]{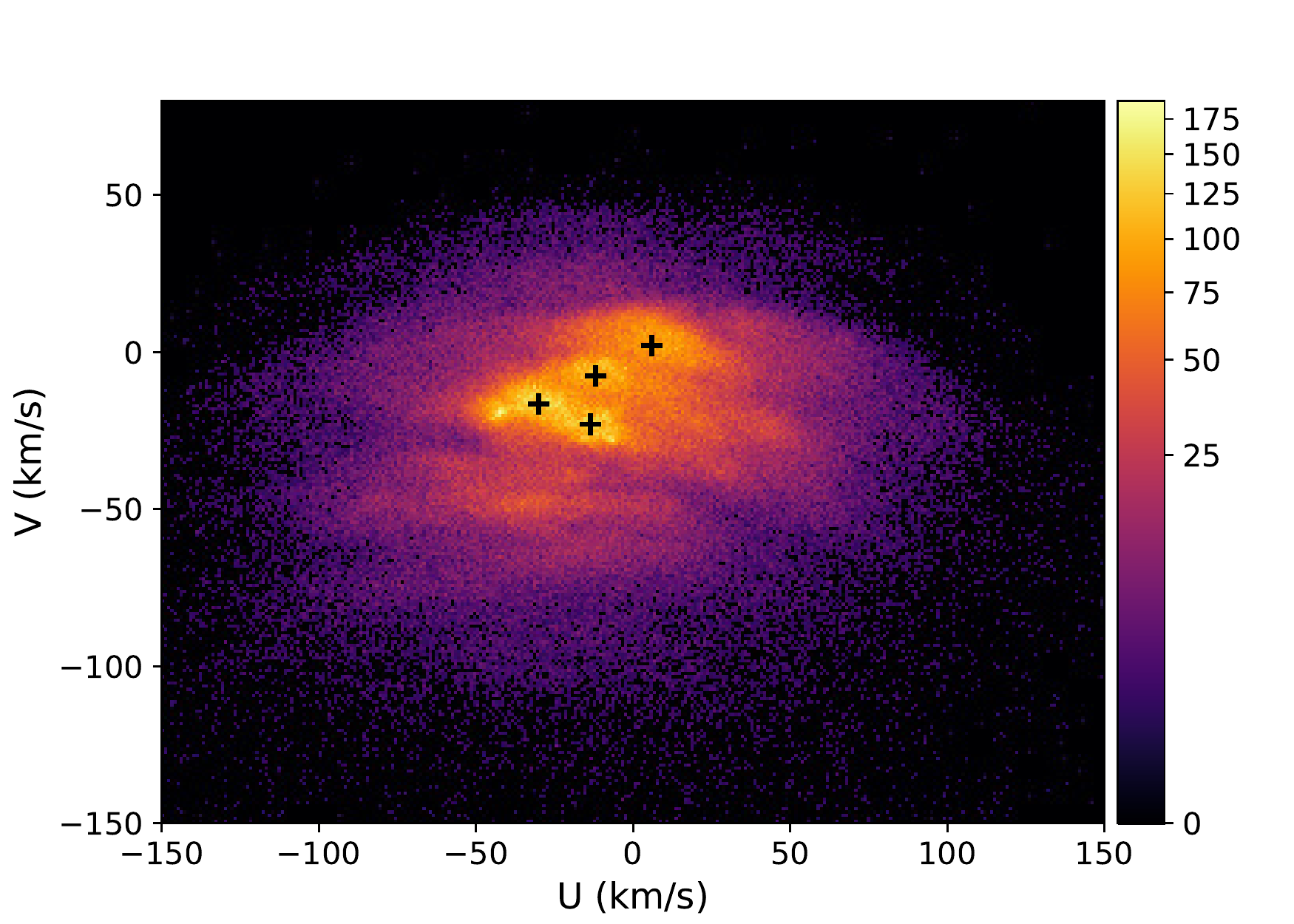}
\caption{U-V velocities of stars within a 200 pc-radius
 solar neighbourhood listed in $\textit{Gaia}$ DR2. The 2D histogram
 has a size of 1 $\kms$. The colour scale shows the star number per
 ($\kms$)$^2$. The plus signs show approximately the positions of the
 four main density maxima.} 
\label{figura11}
\end{figure}

The local U-V velocity field with respect to the Sun resulting from
stable resonant orbital
families must be compared with the observed field. Around zero
velocity, this observed U-V field is shown in Fig.~\ref{figura11}
taking stars within a 200 pc-radius solar neighbourhood 
listed in $\textit{Gaia}$ DR2 \citep{2018A&A...616A...1G}.
 The three main branches
identified by \citet{1999MNRAS.308..731S} appear around the centre,
moving downwards in the V-axis: the Sirius branch, the Middle branch
or Coma Berenices branch, and the Hyades-Pleiades branch.
Also, the Hercules branch at V $\approx$ $-50$ $\kms$ and
other structures below this level. 

The resulting
U-V contributions of periodic orbits crossing the solar
vicinity, in all the stable sections of all the resonant families,
are given in Figs.~\ref{figura1ap} -- \ref{figura10ap}. 
Each figure has its corresponding value of $\Omega_{\rm b}$, and 
the angle $\phi_B$ appears in each panel. These results were obtained
within the 500 pc-radius solar vicinity, as stated above. Each colour
gives the family, as in Fig.~\ref{figura2}. In these figures
we show with plus signs the four main density maxima 
in Fig.~\ref{figura11}. We stress that the
agglomerations obtained in these figures represent only contributions
to the observed local U-V field that could result from orbital
trapping by some resonant families, and do \textit{not} represent
the total field as shown in Fig.~\ref{figura11}, which includes 
possibly trapped and non-trapped stars.

Some relevant points from these results are the following:

\begin{enumerate}

\item With $\Omega_{\rm b}$ $\approx$ 35 -- 40 $\kmskpc$ the main
contributions in U-V velocities due to orbital trapping come from
resonant families 4,\,5,\,6,\,I.
At the other extreme $\Omega_{\rm b}$ $\approx$ 50 -- 57.5 $\kmskpc$,
families IV,VI,VII give the main contributions. This reflects the 
downwards shift of resonant families shown in
Figs.~\ref{figura2},~\ref{figura3},~\ref{figura4}:
the solar vicinity stays at position $R_0$, while resonant families
decrease their distance to the Galactic centre as
$\Omega_{\rm b}$ increases.

\item Some aglomerations near zero velocity, like those observed in
Fig.~\ref{figura11}, do not appear with 
$\Omega_{\rm b}$=42.5,\,45,\,47.5 $\kmskpc$. In our attempt to find
a connection between resonant families and the observed U-V field,
these values of $\Omega_{\rm b}$ are not considered for further
analysis. Thus, the effect of orbital trapping in the local
U-V field is expected to be significant only in the low and high
values of $\Omega_{\rm b}$.
  
\item With $\Omega_{\rm b}$=35 -- 40 $\kmskpc$ there are some
contributions of family VII, in red colour, at high values in
velocity V with $\phi_B$=40$\grad$, 45$\grad$. This result 
is commented in the following section. 

\item In the cases with low angular velocity, 
$\Omega_{\rm b}$=35,\,37.5,\,40 $\kmskpc$, there are some dispersed
contributions with V velocities mainly shifted below zero. An extended
analysis can be made with $\Omega_{\rm b}$=37.5,\,40 $\kmskpc$
taking the angles $\phi_B$=20$\grad$ -- 40$\grad$. 

\item The cases $\Omega_{\rm b}$=50 -- 57.5 $\kmskpc$
give contributions of three families IV,VI,VII near
zero velocity, whose possible relation with the three main branches
identified by \citet{1999MNRAS.308..731S} can be further analysed.

\end{enumerate}

In the next section we proceed with the analysis of some cases in
points (iv) and (v).

\section{The local U-V velocity field of trapped orbits} 
\label{uvatrap}

\subsection{Trapped Orbits by Periodic Orbits crossing the Solar
neighbourhood}
\label{uvatrap1}

The agglomerations in the U-V plane obtained in the last section
need a further analysis, taking into account the possible orbital
trapping of the corresponding periodic orbits.
To represent trapped orbits around stable periodic orbits in a given
family a variation of $V_{x^{\prime}}$ was considered from its value
$V_{x^{\prime}}$=0 at its perpendicular crossing with the 
$x^{\prime}$-axis, keeping the original value of the Jacobi constant
$E_{\rm J}$. Thus, we move in the vertical direction at the
$x^{\prime}$ position of the periodic orbit in an island region of
a Poincar\'e diagram, as in Fig.~\ref{figura5}, but staying within 
this region.

 The trapped orbits will be in general 3D orbits, and
will have projected
tube orbits on the $x^{\prime}$,$y^{\prime}$ plane whose velocity
($V_{x^{\prime}}$,$V_{y^{\prime}}$) can be represented with planar tube
orbits around stable resonant families on the Galactic plane
\citep{2015MNRAS.451..705M}.
The maximum value of $V_{x^{\prime}}$ depends on the given family;
we found approximately a representative top limit around 20 $\kms$ in
the families analysed in this section (but in particular for the
value of $E_{\rm J}$ in panel a of 
Fig.~\ref{figura5} this limit is around 40 $\kms$; thus, the
top limit in $V_{x^{\prime}}$ will be changed in some families), and considered the
discrete values $V_{x^{\prime}}$=5,\,10,\,15,\,20 $\kms$ to initiate
tube orbits. These orbits were computed up to a total time of
10$t_{\rm per}$, with $t_{\rm per}$ the time in which the associated
periodic orbit closes in the non-inertial
($x^{\prime}$,$y^{\prime}$) reference frame. In most cases this total
time is around 5 Gyr. 

Given $\Omega_{\rm b}$, $\phi_B$ and a periodic orbit in a stable
section of a resonant family, with its computed tube orbits and
their crossings with the corresponding 500 pc-radius solar vicinity,
the velocities U,V in Eqs.~\ref{u},~\ref{v} were obtained at some inner
division points. All these points (U,V) in tube orbits are associated
with the periodic orbit. All the periodic orbits in a stable section of
the resonant family will contribute with a different number of points
(U,V). To estimate the relative number contribution of each periodic
orbit, a random sample of 5 $\times$ 10$^4$ stars in  
$\textit{Gaia}$ DR2 \citep{2018A&A...616A...1G} within the 
500 pc-radius solar vicinity
was considered in a characteristic energy versus $E_{\rm J}$ diagram,
 like in
Figs.~\ref{figura2},~\ref{figura3},~\ref{figura4}. The number of points
in this diagram along a thin band around the stable section of the
given resonant family was counted for each $E_{\rm J}$ in the section,
taking the lenght $\Delta E_{\rm J}$ as the mean separation of periodic orbits in
our computations. These numbers give an estimate of the relative
number of trapped orbits along the stable section of the family, and
they were employed to rescale the numbers of computed points (U,V) in
tube orbits at each $E_{\rm J}$.

\begin{figure}
\includegraphics[width=\columnwidth]{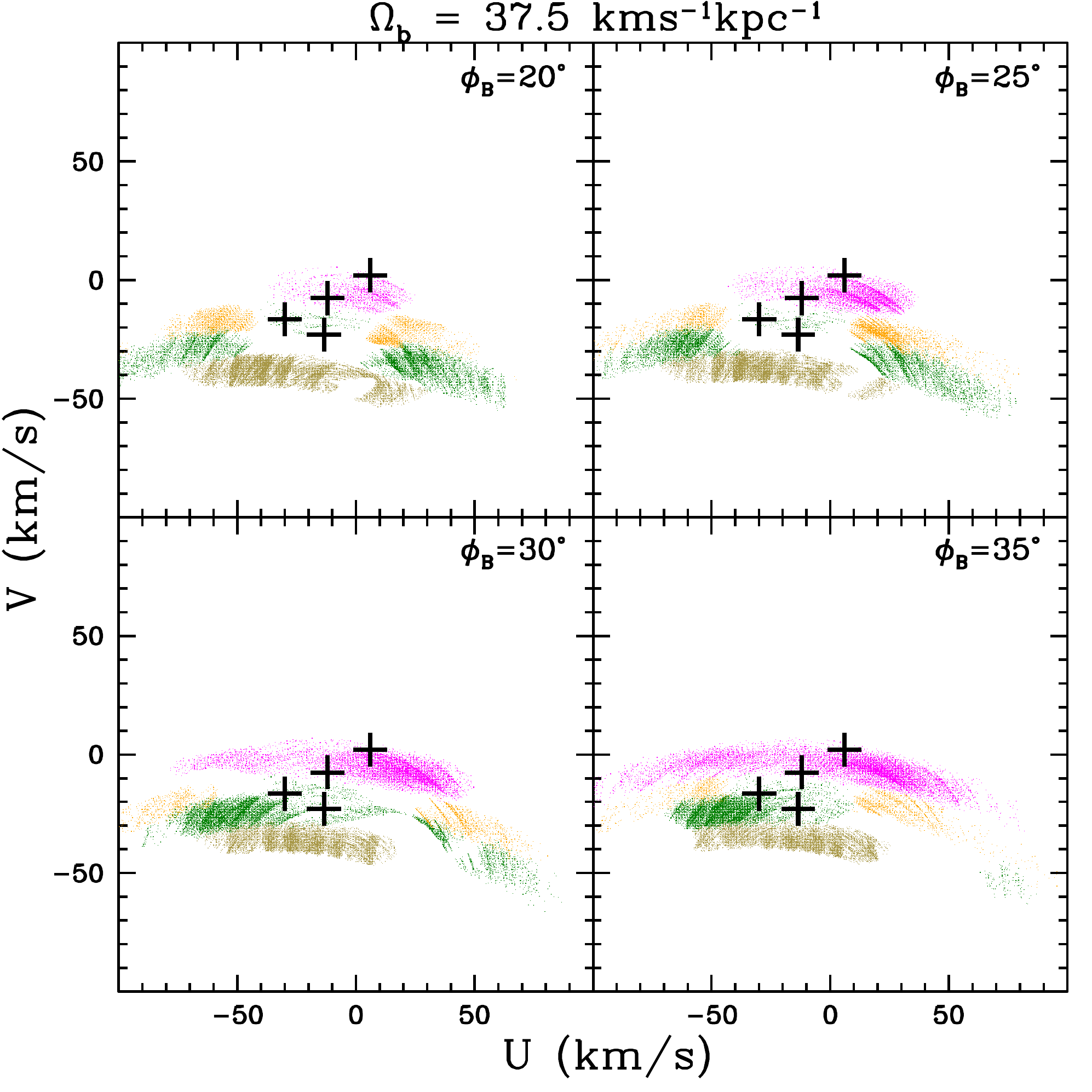}
\caption{Resonant bands produced by families 4,\,5,\,6,\,I, taking
tube orbits around periodic orbits in stable family sections with
$\Omega_{\rm b}$=37.5 $\kmskpc$. See the family colour in
Fig.~\ref{figura2}. The value of the angle $\phi_B$ appears in each
panel. In order to compare with Fig.~\ref{figura11} here the solar
vicinity has a radius of 200 pc. We show with plus signs the four
main density maxima in Fig.~\ref{figura11}.}
\label{figura12}
\end{figure}

\begin{figure}
\includegraphics[width=\columnwidth]{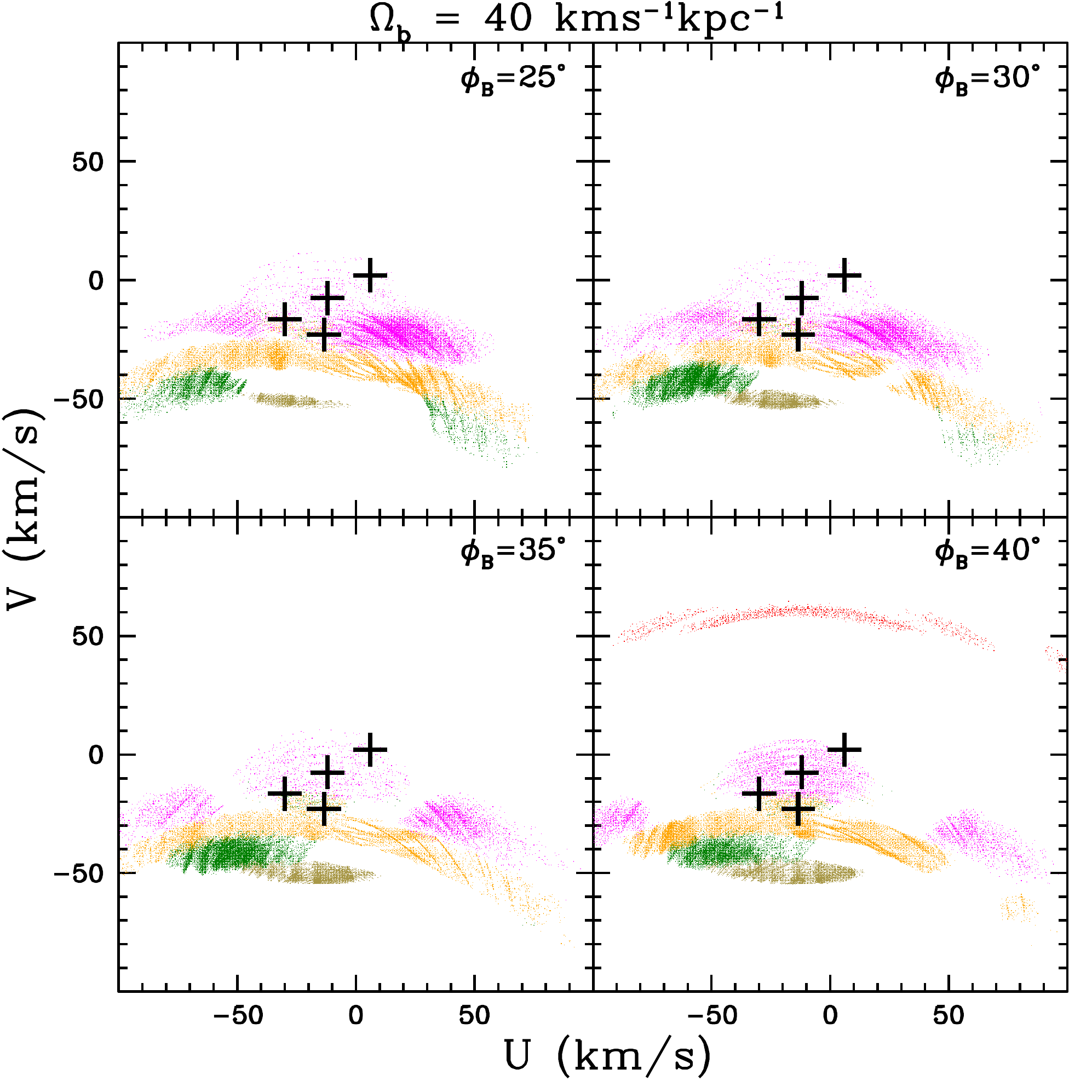}
\caption{Same as Fig.~\ref{figura13}, here with
$\Omega_{\rm b}$=40 $\kmskpc$. Note the Hercules branch at
 V=$-50$ $\kms$ produced by family 4, or resonance 8/1.}
\label{figura13}
\end{figure}

\begin{figure}
\includegraphics[width=\columnwidth]{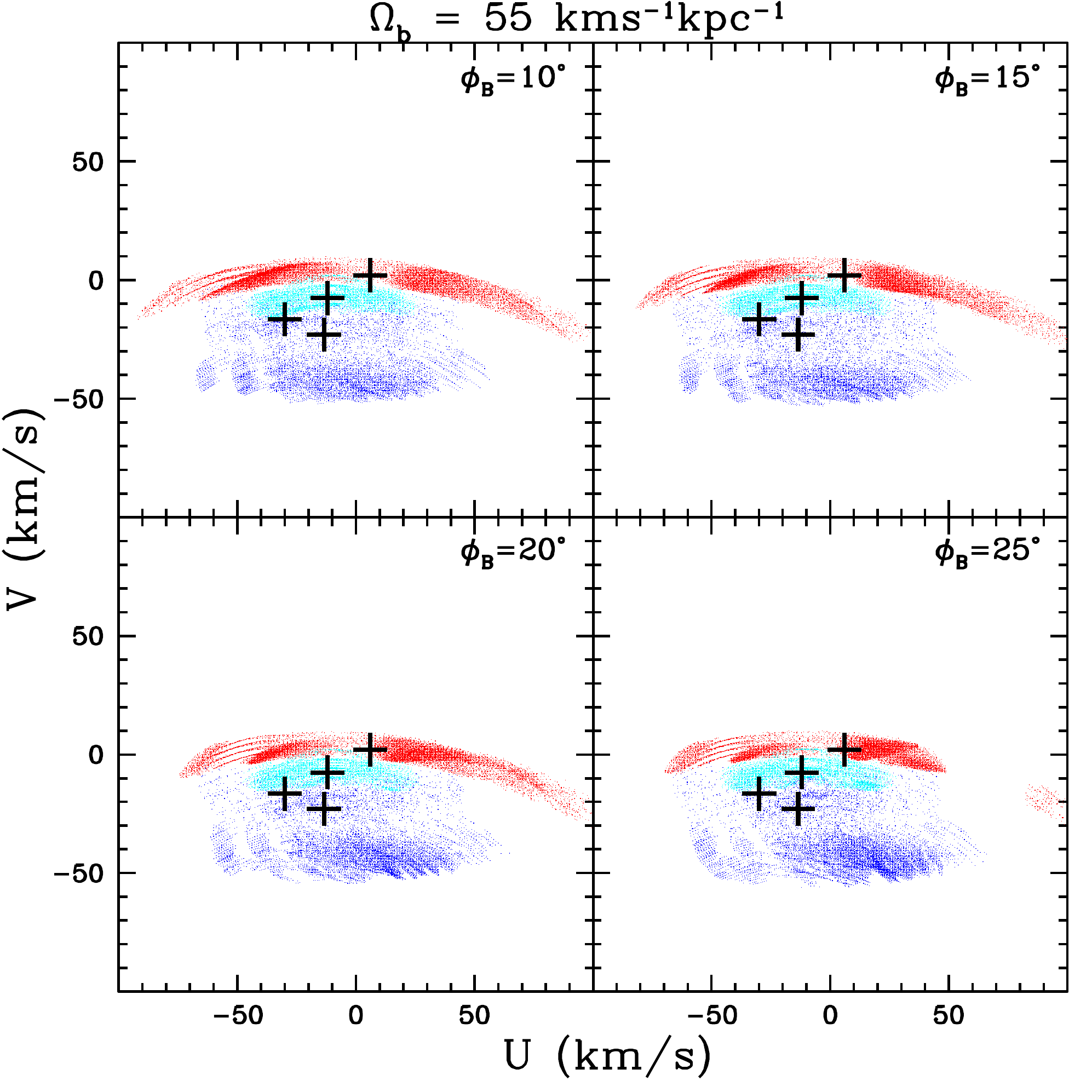}
\caption{As in Fig.~\ref{figura13}, with
$\Omega_{\rm b}$=55 $\kmskpc$; here the resonant bands are
generated by families IV,VI,VII.}
\label{figura14}
\end{figure}

\begin{figure}
\includegraphics[width=\columnwidth]{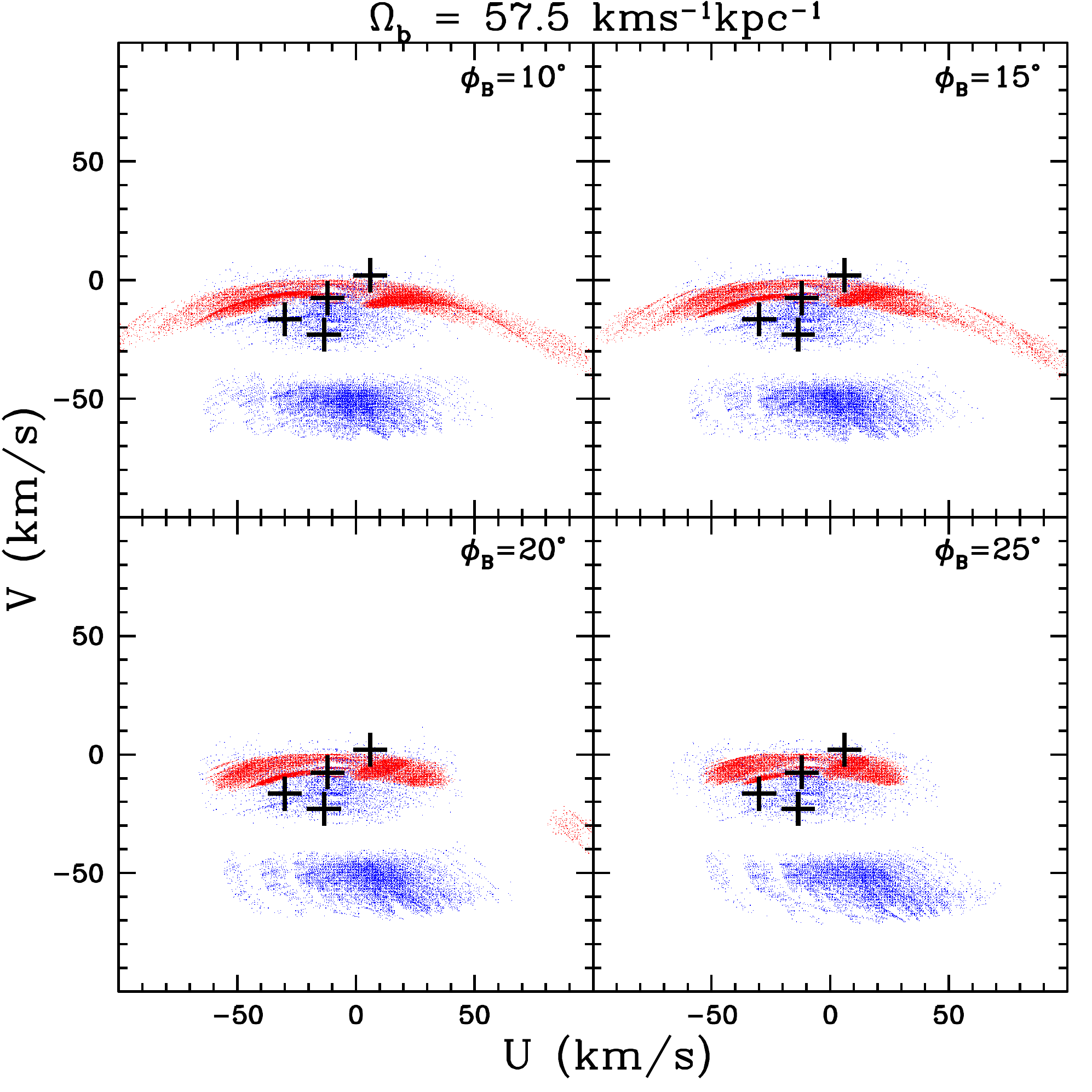}
\caption{As in Fig.~\ref{figura13}, with
$\Omega_{\rm b}$=57.5 $\kmskpc$; here the resonant bands are
generated by families IV and VII.}
\label{figura15}
\end{figure}

We followed this procedure to analize the cases
with $\Omega_{\rm b}$=37.5,\,40 $\kmskpc$ in 
Figs.~\ref{figura2ap},~\ref{figura3ap}, taking the angles
$\phi_B$=20$\grad$ -- 35$\grad$ in the first case, and 
$\phi_B$=25$\grad$ -- 40$\grad$ in the second. Also the cases
$\Omega_{\rm b}$=55,\,57.5 $\kmskpc$ in
Figs.~\ref{figura9ap},~\ref{figura10ap} with
$\phi_B$=10$\grad$ -- 25$\grad$.
The results are shown in
Figs.~\ref{figura12}--\ref{figura15}, taking points (U,V) only within a
200 pc-radius solar vicinity, in order to compare with
Fig.~\ref{figura11}.

\begin{figure}
\includegraphics[width=\columnwidth]{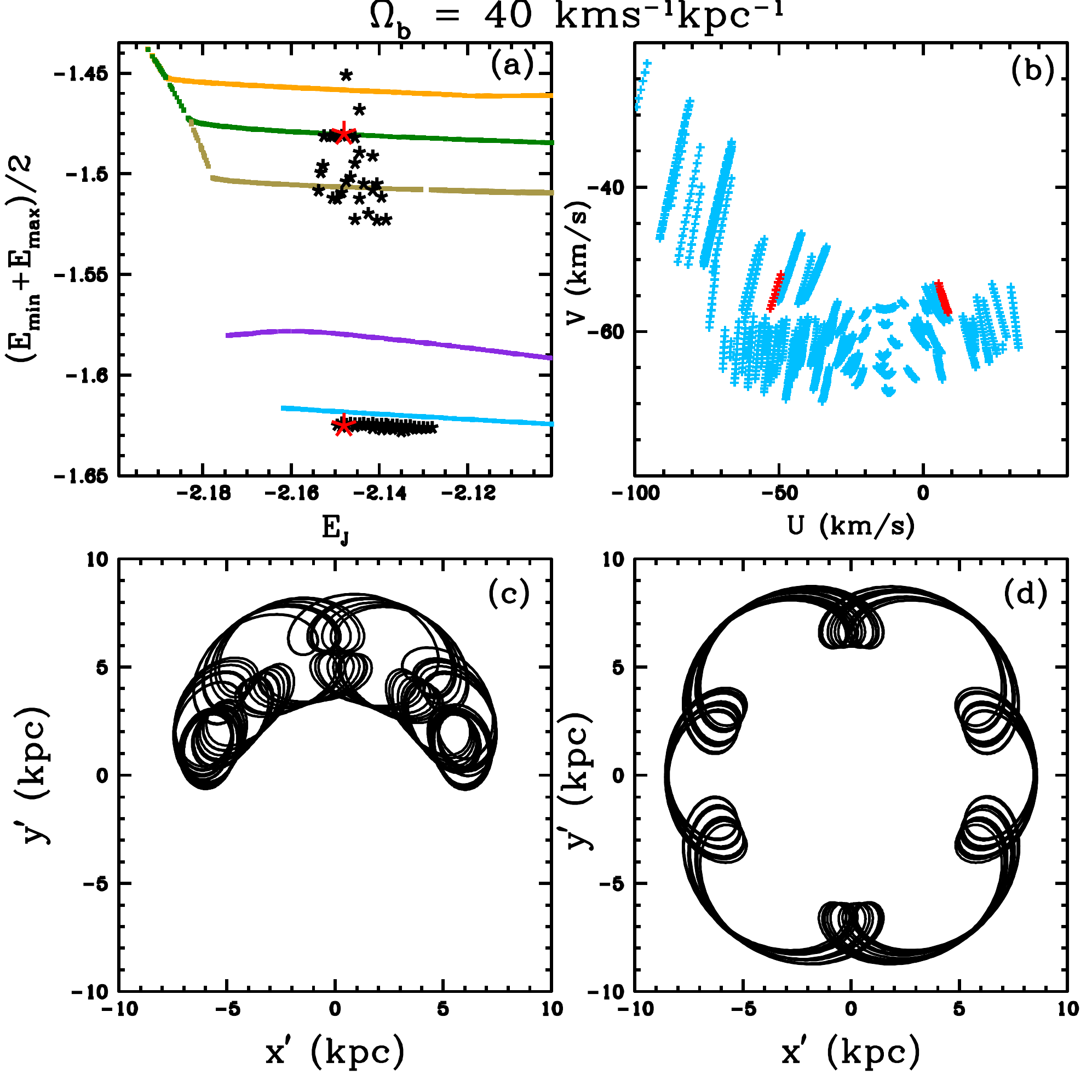}
\caption{Orbits constructed from family 2 can give points U,V
around the Hercules stream. Here $\Omega_{\rm b}$=40 $\kmskpc$, 
$\phi_B$=40$\grad$ are considered, with a 500 pc-radius solar vicinity.
Panel (a) shows a region of the characteristic energy vs $E_J$ diagram
with families 2,3,4,5,6. The black starred points near the blue curve 
of family 2 represent orbits inside the island regions around some of
its periodic orbits but close to their boundaries. The black points
lying in the region of families 4,5,6 (brown, green, and orange curves)
represent orbits taken outside the island regions of family 2.
The red points are two particular orbits. Panel (b) shows the resulting
U-V plane due to the starred points in panel (a). The (U,V) points
distribute around the Hercules stream. Panels (c) and (d) show the
orbits corresponding to the lower and upper red points in panel (a).}
\label{figura16}
\end{figure}

In each case shown in these figures the resonant families generate
some bands of points U,V. Up to four bands can appear in the low values
$\Omega_{\rm b}$=37.5,\,40 $\kmskpc$, resulting from
families 4,\,5,\,6,\,I.  Three main bands are obtained with
families IV,VI,VII in the other extreme
$\Omega_{\rm b}$=55,\,57.5 $\kmskpc$. The inclination of these bands
with respect to the U-axis is slightly greater with low
$\Omega_{\rm b}$, and their width is similar to the
width of the three main branches in Fig.~\ref{figura11}, the Sirius,
Coma Berenices, and Hyades-Pleiades branches. 

The cases with $\Omega_{\rm b}$=40,\,55,\,57.5 $\kmskpc$
give structures around level V=$-50$ $\kms$, close to the Hercules
branch. The case $\Omega_{\rm b}$=40 $\kmskpc$ approximates better
the agglomeration of bands around this V-level shown in the 
\textit{Gaia} data in Fig.~\ref{figura11}. In this case the Hercules
branch is produced by family 4 (brown colour), or resonance 8/1 outside
corotation, and the near families 5 (dark green colour) and 6 (orange
colour), i.e. respectively resonances 6/1 and 5/1, contribute above
family 4. 

Another interesting structure which can be approximated with
$\Omega_{\rm b}$=40 $\kmskpc$, and specially with $\phi_B$=40$\grad$,
is the newly detected low-density arch at V $\simeq$ 40 $\kms$, which
has been commented by \citet{2018A&A...616A..11G} in the U-V plane
of their figure 22, and appears in our similar Fig.~\ref{figura11}.
 In our results
this arch is approximated by orbital trapping due to family VII, i.e.
the resonance 4/3.

In the dynamical model of 
\citet{2017ApJ...840L...2P} with $\Omega_{\rm b}$=39 $\kmskpc$
the Hercules stream is approximated by stars orbiting the Lagrange
points $L_4$, $L_5$, thus associated with our family 2
(see Fig.~\ref{figura6}), or circulating between both points.
Points U,V due to family 2 do not appear in
Figs.~\ref{figura12},~\ref{figura13} because for the computations in
this section, which refer to orbital trapping due to \textit{periodic
orbits crossing the solar neighbourhood}, we did not find periodic
orbits of family 2 satisfying this condition. In the next section we
consider the missing contribution in the solar vicinity of tube orbits
around periodic orbits in family 2, and other families, which are
external to this vicinity.

To show how orbits constructed from family 2 can give points 
U,V around the Hercules stream, a similar computation to that made by
\citet{2017ApJ...840L...2P} is shown in Fig.~\ref{figura16},
using $\Omega_{\rm b}$=40 $\kmskpc$ with $\phi_B$=40$\grad$.
Panel (a) in this figure is the characteristic energy vs $E_J$
diagram showing only a region with families 2,3,4,5,6. The black 
starred points near the blue curve of family 2 represent orbits  
inside the island regions around some of its periodic orbits but close
to their boundaries. The black points lying in the region of
families 4,5,6 (brown, green, and orange curves) represent orbits taken
outside the island regions of family 2, i.e. they are not necessarily
circulating between the Lagrange points $L_4$, $L_5$, but have jumped
to another region. The red points are two  
particular orbits in these two sets. Panel (b) shows the resulting U-V
plane due to all black and red points in panel (a), taking a
500 pc-radius solar vicinity. The (U,V) points distribute around the
Hercules stream. Thus, points in this region can be obtained by
orbital trapping in the corotation resonance and close resonant
families. Panels (c) and (d) show, respectively, the orbits
corresponding to the lower and upper red points in panel (a).
An important continuation of this Fig.~\ref{figura16} is presented
in the following section, taking values of $E_J$ greater than those
considered in panel (a).

\subsection{Trapped Orbits by Periodic Orbits not crossing the Solar
neighbourhood}
\label{uvatrap2}

Now we take into account orbital trapping by stable sections of
resonant families whose \textit{periodic} orbits do not cross the given
solar vicinity. In these cases, the amplitude of the corresponding tube
orbits around periodic orbits is increased until they cross the solar
vicinity, and their velocity field (U,V) is computed following the
procedure given in Section~\ref{uvatrap1}. In the present section
only two cases are considered of those analysed in
Section~\ref{uvatrap1}: $\Omega_{\rm b}$=40 $\kmskpc$ with 
$\phi_B$=40$\grad$ from Fig.~\ref{figura13}, and
$\Omega_{\rm b}$=55 $\kmskpc$ with $\phi_B$=10$\grad$ from
Fig.~\ref{figura14}.

With $\Omega_{\rm b}$=40 $\kmskpc$ the additional resonant
families which are analysed are 2,3,IV,V,VI,VIII; with
$\Omega_{\rm b}$=55 $\kmskpc$ the additional families are
2,3,V,VIII,IX. The contributions of all these families are plotted only
in the interval [-100,100] $\kms$ of velocities U,V, as in figures of
the preceeding section.  

\begin{figure}
\includegraphics[width=\columnwidth]{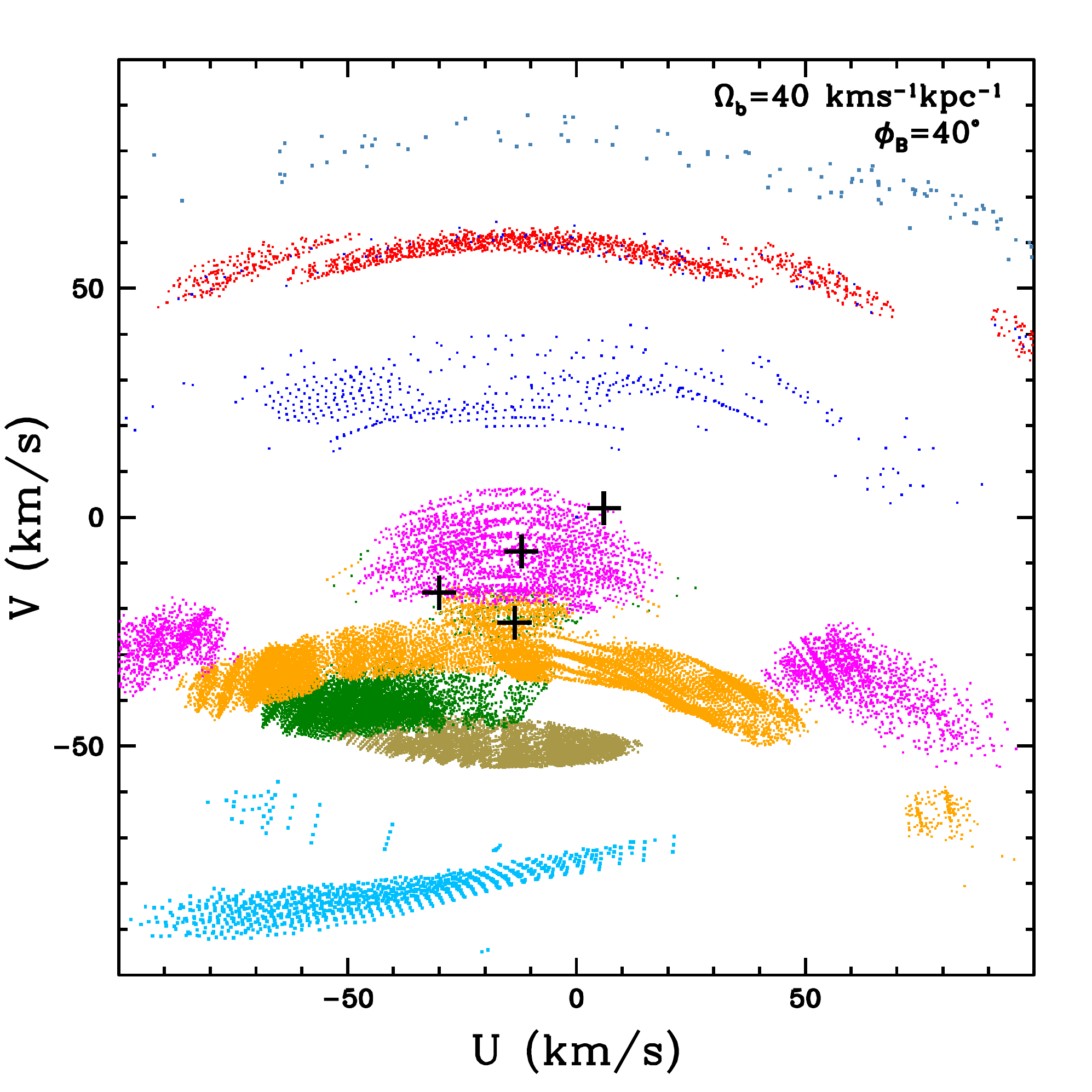}
\caption{The complete case
$\Omega_{\rm b}$=40 $\kmskpc$, $\phi_B$=40$\grad$, adding the
contributions obtained in Section~\ref{uvatrap2} to those shown in
Fig.~\ref{figura13}. The few steel blue (see family colours in
Fig.~\ref{figura6}) points in the arch at
V $\simeq$ 80 $\kms$ correspond to family VIII, resonance 1/1.
Points like these are not detected in Fig.~\ref{figura11}.
The blue points in the arch between 20 -- 40 $\kms$ in velocity V
correspond to family IV, resonance 2/1. Points like these do appear in
Fig.~\ref{figura11}. The sky blue points in the inclined structure
below the Hercules branch correspond to family 2, corotation
resonance. This structure is detected in Fig.~\ref{figura11}.}
\label{figura17}
\end{figure}

In Fig.~\ref{figura17} we show the complete case 
$\Omega_{\rm b}$=40 $\kmskpc$, $\phi_B$=40$\grad$, adding the
contributions obtained in this section to those shown in
Fig.~\ref{figura13}. The few steel blue (see family colours in
Fig.~\ref{figura6}) points in the arch at 
V $\simeq$ 80 $\kms$ correspond to family VIII, resonance 1/1.
Points like these are not detected in Fig.~\ref{figura11}.
The blue points in the arch between 20 -- 40 $\kms$ in velocity V
correspond to family IV, resonance 2/1. Points like these do appear in
Fig.~\ref{figura11}. The sky blue points in the inclined structure
below the Hercules branch correspond to family 2, corotation
resonance. This structure is detected in Fig.~\ref{figura11}.

\begin{figure}
\includegraphics[width=\columnwidth]{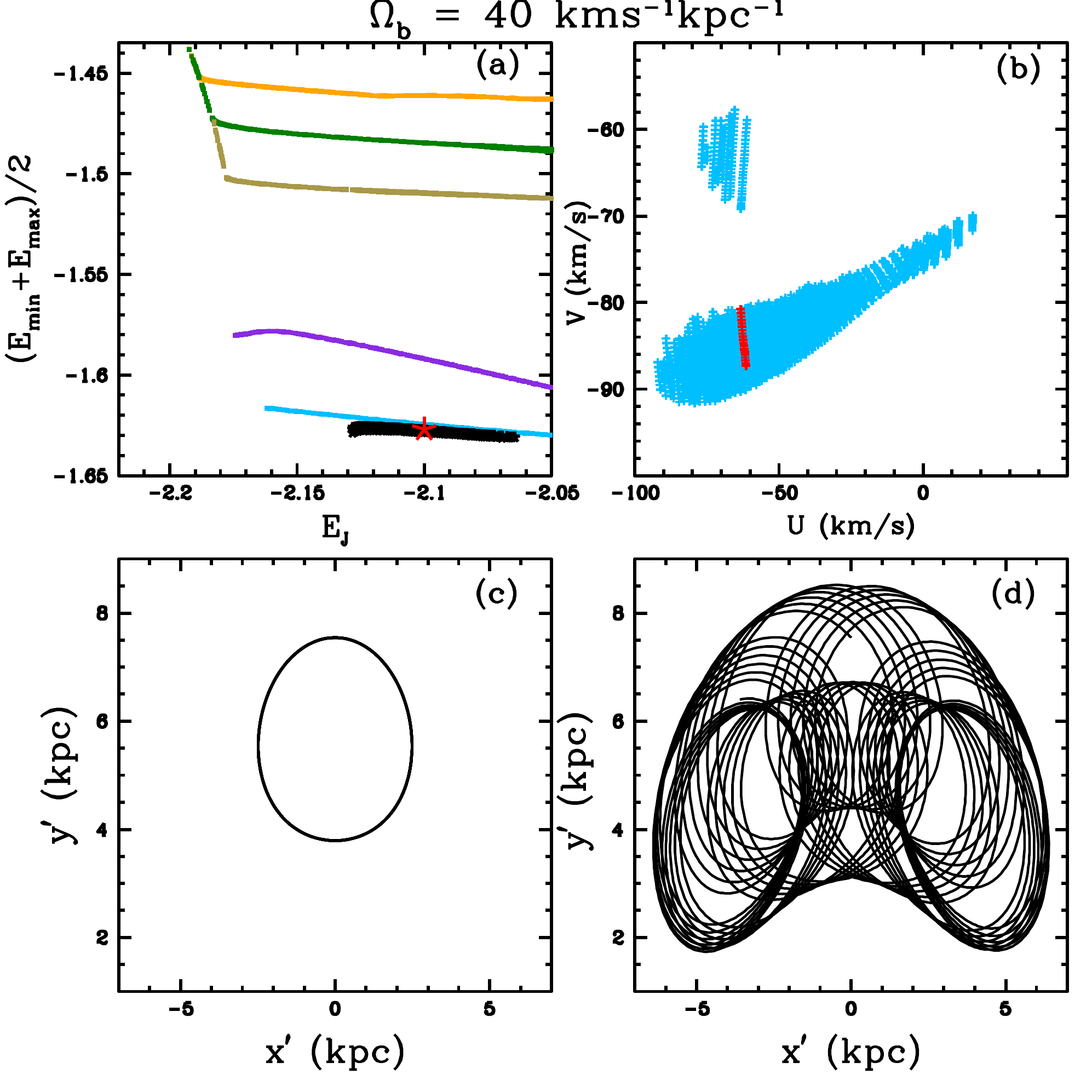}
\caption{Analysis of the inclined structure due to family 2 in
Fig.~\ref{figura17}. Panel (a) is a continuation of panel (a) in
Fig.~\ref{figura16}, here taking greater values of $E_J$. The dense
black points close to the blue curve of family 2 represent tube
orbits trapped by periodic orbits in this family, with their
amplitudes taken such that they may cross the solar vicinity.
The red point is a particular orbit. Panel (b) shows the velocity
points (U,V) generated by all these orbits within the solar vicinity;
the red points correspond to the red point in panel (a). Thus, these
orbits generate the inclined structure in Fig.~\ref{figura17}.
Panel (c) shows the stable periodic orbit in family 2 at the value
$E_J$ of the red point in panel (a), and panel (d) is the tube orbit
of this red point.}
\label{figura18}
\end{figure}

In Fig.~\ref{figura18} the inclined structure due to family 2
is further analysed. Panel (a) is a continuation of panel (a) in
Fig.~\ref{figura16}, here taking greater values of $E_J$. The dense
black points close to the blue curve of family 2 represent tube
orbits trapped by periodic orbits in this family, with their
amplitudes taken such that they may cross the solar vicinity.
The red point is a particular orbit. Panel (b) shows the velocity
points (U,V) generated by all these orbits within the solar vicinity;
the red points correspond to the red point in panel (a). Thus, these
orbits generate the inclined structure in Fig.~\ref{figura17}.
Panel (c) shows the stable periodic orbit in family 2 at the value
$E_J$ of the red point in panel (a), and panel (d) is the tube orbit
of this red point.

\begin{figure}
\includegraphics[width=\columnwidth]{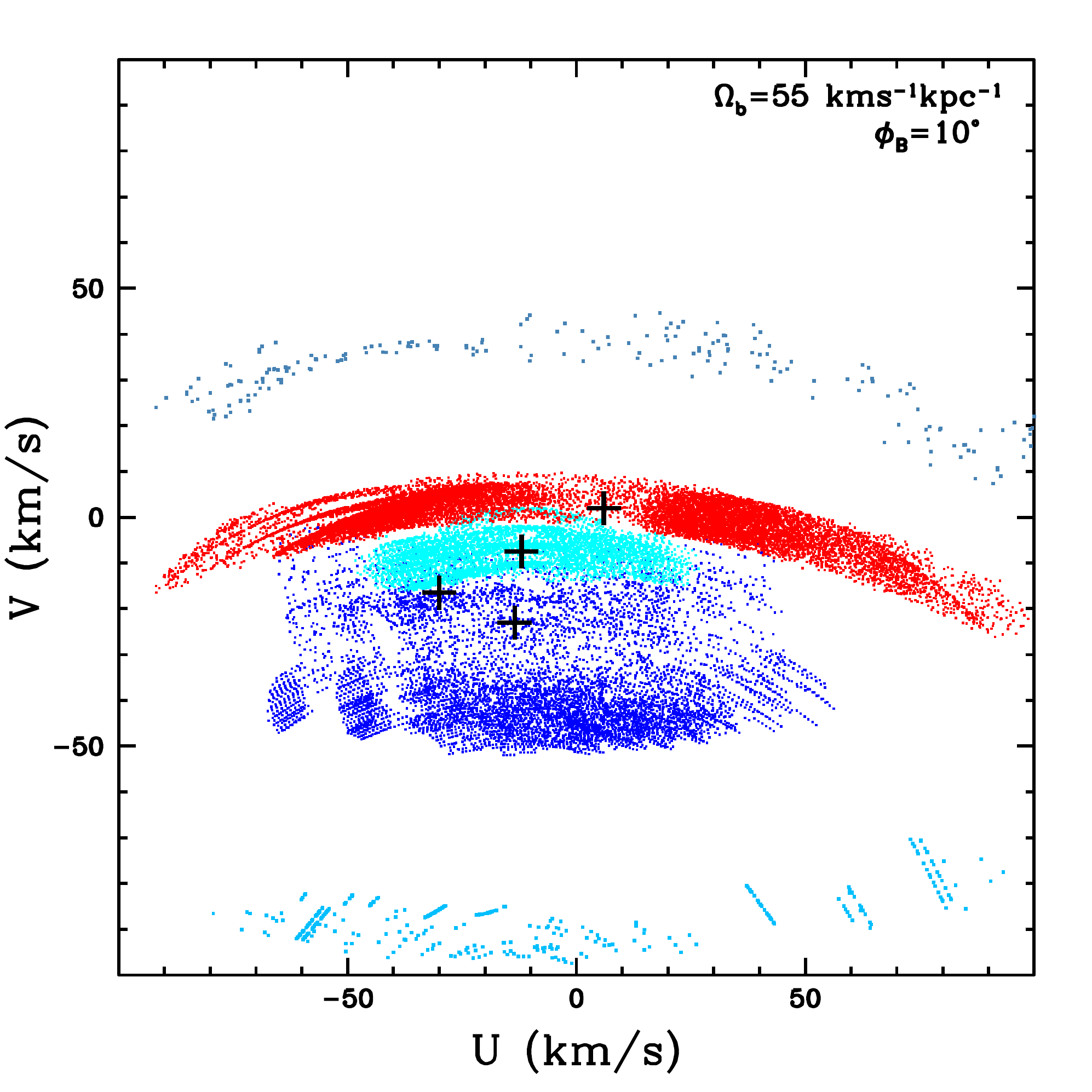}
\caption{The complete case
$\Omega_{\rm b}$=55 $\kmskpc$, $\phi_B$=10$\grad$, adding the
contributions obtained in Section~\ref{uvatrap2} to those shown in
Fig.~\ref{figura14}. Family VIII contributes with the arch at
30 -- 40 $\kms$ in velocity V, and family 2 gives the points
around V $\simeq$ $-90$ $\kms$.}
\label{figura19}
\end{figure}

In Fig.~\ref{figura19} we show the complete case
$\Omega_{\rm b}$=55 $\kmskpc$, $\phi_B$=10$\grad$, adding the
contributions obtained in this section to those shown in
Fig.~\ref{figura14}. Family VIII contributes with the arch at
30 -- 40 $\kms$ in velocity V, and family 2 gives the points
around V $\simeq$ $-90$ $\kms$. In this case the inclined feature
in Fig.~\ref{figura17} is not obtained.

With these results, our
best approximation to the local velocity field by orbital trapping
is $\Omega_{\rm b}$=40 $\kmskpc$ with $\phi_B$=40$\grad$.
Based on data taken by
APOGEE-2S, the analysis given by \citet{2018MNRAS.474...95H}
favor a fast bar with $\Omega_{\rm b}$=48.6 $\kmskpc$. More
recent studies support a slow bar \citep{2019A&A...626A..41M,
2019A&A...632A.107M,2019MNRAS.488.4552S,2019MNRAS.489.3519C,
2019MNRAS.490.4740B} with $\Omega_{\rm b}$ around 40 $\kmskpc$.
In particular, \citet{2019A&A...626A..41M} show
that a bar with $\Omega_{\rm b}$=39 $\kmskpc$ creates most of the
observed structures in local action space which are related to
resonances generated by the bar. We identify some bands of points
U,V in Fig.~\ref{figura17} with resonance lines in their figure 4:
our families I and 5 are related with their resonances 4/1 and 6/1,
respectively; the band of family 6 is the resonance 5/1 not shown
in their figure 4; the resonance 3/1, associated to family III, is
slightly appearing with $\Omega_{\rm b}$=40 $\kmskpc$,
$\phi_B$=10$\grad$,15$\grad$,20$\grad$ in Fig.~\ref{figura3ap}, but
is strong with $\Omega_{\rm b}$=42.5,\,45 $\kmskpc$ in
Figs.~\ref{figura4ap},~\ref{figura5ap}. The low-density arch
at V $\simeq$ 40 $\kms$ in Fig.~\ref{figura11} appears to be related
with the outer Lindblad resonance 2/1 in figure 4 of
\citet{2019A&A...626A..41M}, and is obtained by orbital simulations 
with a slow bar given by \citet{2018MNRAS.477.3945H}.
With our results in Fig.~\ref{figura17} this arch can be
approximated with family VII, i.e. resonance 4/3, but also the wide
arch due to resonance 2/1 (family IV) is close to this structure.

Our analysis has considered only the effects of bar resonances
in the U-V local velocity field, but there are other studies which
include also a spiral pattern. With test particle simulations,
\citet{2007A&A...467..145C} concludes that a bar,
with an angular velocity of $\Omega_{\rm b}$=57.4 $\kmskpc$, and
a 4-armed spiral can approximate the main concentrations in the U-V
plane. A similar conclusion is reached by \citet{2018A&A...615A..10M,
2018ApJ...863L..37M}, with a 4-arms spiral and a bar with low mass
$10^9 M_{\odot}$, both rotating with an angular velocity
28.5 $\kmskpc$. \citet{2011MNRAS.418.1423A}, also with test particle
simulations, including a bar and a 2-armed spiral, obtain concentrations
similar to the observed, but the appearence of these features depends
on the integration time. Regarding the Hercules branch,
\citet{2018A&A...615A..10M,2018ApJ...863L..37M} find that this
feature emerges in the 8/1 resonance due to the spiral arms, and the
other observed main concentrations are related with the corotation
zone of these arms. This result with their low mass
$10^9 M_{\odot}$ of the bar, is different from the one we find here
with a bar of $10^{10} M_{\odot}$ and no spiral arms: the Hercules
branch is mainly generated by the 8/1 resonance of this bar.
The model of \citet{2018A&A...615A..10M,2018ApJ...863L..37M} is
supported by a recent study of \citet{2020ApJ...888...75B}, who show
that with that model the local structures in the U-V velocity
distribution are well reproduced.
Thus, it would be interesting to pursue
an analysis to see under what conditions the approximate matches we have
obtained with a bar of $10^{10} M_{\odot}$ can be sustained including
a spiral pattern, and how well resonant families due to the bar
survive.

\subsection{Trapped Orbits in the Galactic Halo}
\label{uvatrap3}

\begin{figure}
\includegraphics[width=\columnwidth]{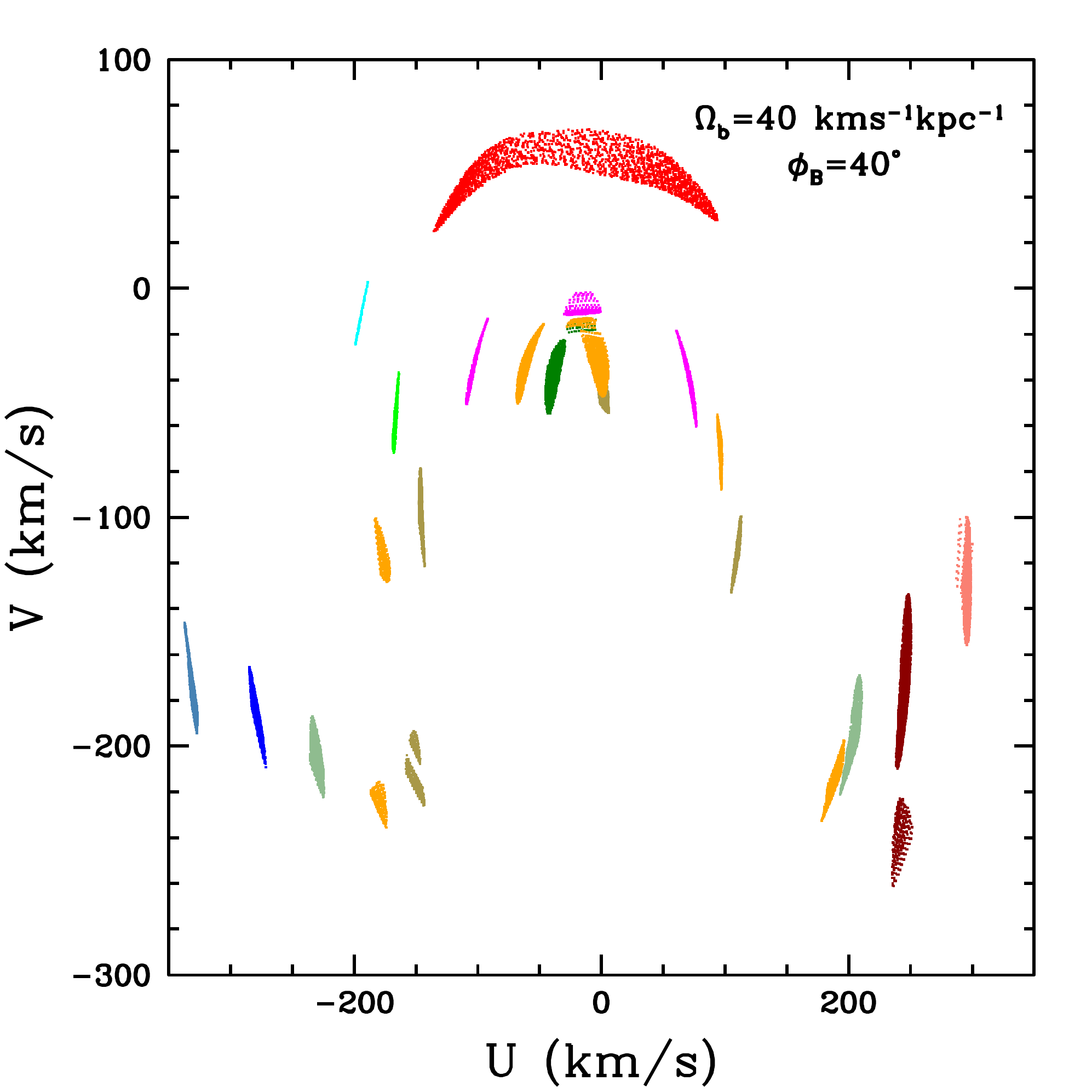}
\caption{Contributions of periodic orbits crossing the 500 pc-radius
solar vicinity in all the analysed resonant families, with
$\Omega_{\rm b}$=40 $\kmskpc$, $\phi_B$=40$\grad$.}
\label{figura20}
\end{figure}

\begin{figure}
\includegraphics[width=\columnwidth]{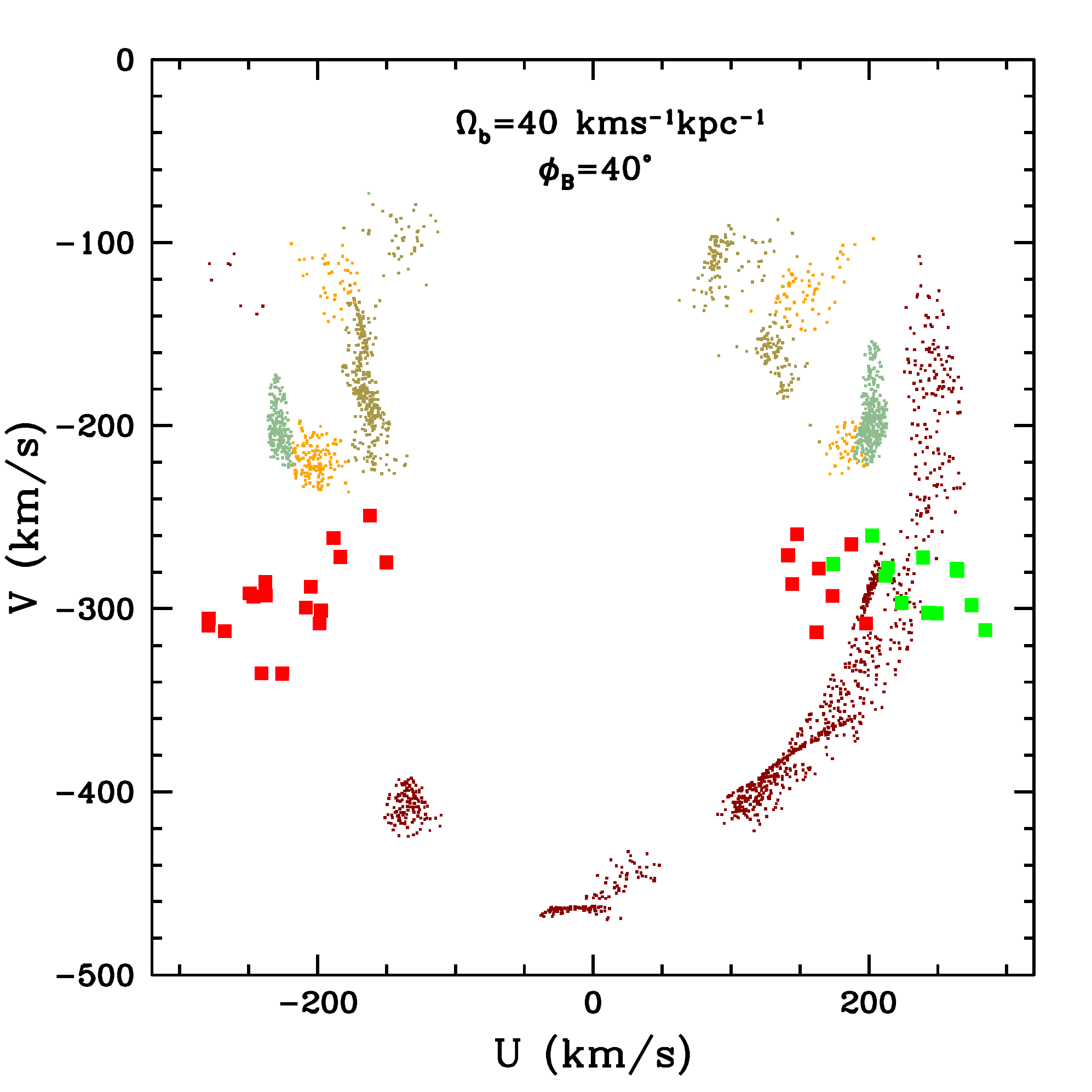}
\caption{Contributions of trapped orbits in regions of
families 6,II,V in Fig.~\ref{figura20}, with family V extended towards
the retrograde region. The points of groups G18-39 and G21-22 in the
Galactic halo are shown with red and green colours.}
\label{figura21}
\end{figure}

We have commented the results obtained around the four main branches
in the U,V plane, but it is interesting to give also the predicted
contributions of orbital trapping in other regions, considering
stellar motions representing the Galactic halo. In this part we 
take our best case $\Omega_{\rm b}$=40 $\kmskpc$, $\phi_B$=40$\grad$. 
Fig.~\ref{figura20} shows the expected contributions of stable 
\textit{periodic} orbits in all the analysed families, taking a
500 pc-radius solar vicinity. This figure is an extension of the
lower left panel in Fig.~\ref{figura3ap}. As stated in
 Section~\ref{fam},
only prograde orbits are considered in our analysis, thus contributions
below V $\approx$ $-260$ $\kms$ are missing in this figure.
No support by orbital trapping is obtained around
(U,V)=(0,$-100$) $\kms$, the region of the Arcturus stream
\citep{1971PASP...83..271E,2009IAUS..254..139W,2019A&A...631A..47K}.
Particular strong contributions of families 6,II,V appear in
the halo region, around U=200 $\kms$. These structures seem to be 
related with star groups G18-39 and G21-22 in the Galactic halo
identified by \citet{2012RMxAA..48..109S} and plotted in the U-V
plane in their figure 7. These two groups have star members
mostly in retrograde orbits. Thus, in order to see if there is a
possible relation between results in Fig.~\ref{figura20} with these
groups, we have extended in particular the family V towards the
retrograde region, and have computed tube orbits which can reach the
solar vicinity. Fig.~\ref{figura21} shows this extended contribution
of family V, along with those of families 6 and II. The updated points
of groups G18-39 and G21-22 are shown with red and green colours, as
in figure 7 of \citet{2012RMxAA..48..109S}. A possible relation with
these groups is obatined only in the right side of the figure.
Further analysis would be needed to see if orbital trapping of both
groups might be an alternate explanation to that proposed by
\citet{2012RMxAA..48..109S}, due to the accretion of a dwarf galaxy.

\section{Conclusions}
\label{concl}

Employing a Galactic potential including a bar, we have analysed the
local U-V velocity field and looked for possible effects of orbital
trapping by bar resonances on the Galactic plane. Fifteen resonant
families have been computed considering ten values of the bar's
angular velocity, $\Omega_{\rm b}$, between 35 and 57.5 $\kmskpc$,
taking in each one ten values of the angle between the bar's major
axis and the Sun-Galactic centre line, $\phi_B$, in the range
5$\grad$ -- 50$\grad$.
In each combination ($\Omega_{\rm b}$,$\phi_B$) of these two
parameters, we consider the orbital
crossings within a 500 pc-radius solar vicinity of the stable
periodic orbits in each resonant family. The resulting U-V field of
these stable orbits in each family gives a first estimate of their
possible contribution to the observed U-V field due to orbital
trapping. Considering only combinations ($\Omega_{\rm b}$,$\phi_B$)
whose resulting contributions may be similar to the observed U-V
field, with its four main branches: Sirius, Coma Berenices,
Hyades-Pleiades, and Hercules, we find that the cases which may give
some similarity with these observed features are in the low range
$\Omega_{\rm b}$=37.5,\,40 $\kmskpc$, and in the high range
$\Omega_{\rm b}$=55,\,57.5 $\kmskpc$. In these cases, and for each
stable periodic orbit in a family crossing the solar vicinity, some
representative tube orbits on the Galactic plane have been computed,
keeping their velocity (U,V) at some orbital division points if they
cross this vicinity. These velocity points will represent 3D
stellar orbits trapped by resonant families on the Galactic plane.
The resulting structures in the local U-V plane form resonant bands
appearing at various levels in velocity V. We find that cases
$\Omega_{\rm b}$=40,\,55 $\kmskpc$ show the greatest similarity with
the observed main branches. 
Regarding particular features in the local U-V plane, we find that
with the slow bar velocity $\Omega_{\rm b}$=40 $\kmskpc$ the Hercules
branch at V=$-50$ $\kms$ can be obtained by orbital trapping in
family 4, i.e. the resonance 8/1 outside corotation, and the
close features produced by resonances 5/1 and 6/1.
Also with this velocity and with an angle $\phi_B$=40$\grad$,
orbital trapping is able to reproduce the newly detected low-density
arch at V $\simeq$ 40 $\kms$ shown by $\textit{Gaia}$ DR2 data.
Adding to these results the orbital trapping contributions
of resonant families whose periodic orbits do not cross the given
solar vicinity, we have found the inclined structure below the
Hercules branch (Fig.~\ref{figura17}), which is also observed in the
 $\textit{Gaia}$ DR2
data. This structure is produced by tube orbits around Lagrange point
$L_5$. Another obtained feature is an arch between 20 -- 40 $\kms$ in
velocity V, due to family IV, i.e. resonance 2/1.
Thus, our best approximation to the local velocity field by orbital
trapping is $\Omega_{\rm b}$=40 $\kmskpc$ with $\phi_B$=40$\grad$.
With this solution, we give some predicted contributions due to orbital
trapping in regions of the U-V plane corresponding to the Galactic
halo, which could help to further restrict the value of the angular
velocity of the Galactic bar. In these regions, no support by orbital
trapping is found for the Arcturus stream at V $\approx$ $-100$ $\kms$. 
Some support is found for the halo moving groups G18-39 and G21-22 at
U $\approx$ 200 $\kms$.
Considering the results of our analysis,
compared with existing studies which employ a bar with a low mass and
include a spiral pattern, it would be interesting to reconsider the 
effect of orbital trapping taking into account, in addition to a bar
of higher mass, the effect of the spiral arms. It is important to see
in this case under what conditions the approximate matches that we
have obtained can be sustained including the spiral arms, and how well
resonant families due to the bar survive.

\section*{Acknowledgements}

We thank an anonymous referee for comments and suggestions that
helped to improve this paper. L.C-V thanks the Fondo Nacional de
Financiamiento para la
Ciencia, La Tecnolog\'ia y la innovaci\'on `FRANCISCO JOS\'E DE
CALDAS', MINCIENCIAS, and the VIIS for the economic support
of this research. A.P-V acknowledges the DGAPA-PAPIIT grant IG100319.

\section*{Data Availability}

Data available on request.







\appendix

\section{U-V contributions of stable periodic orbits in all the
resonant families}
\label{ap1}

Here we show the resulting U-V contributions of
\textit{periodic} orbits crossing a 500 pc-radius solar
vicinity, in all the stable sections of all the resonant families.
Each figure has its corresponding value of $\Omega_{\rm b}$, and
the angle $\phi_B$ appears in each panel. The four main density maxima
in Fig.~\ref{figura11} are shown with plus signs.

\begin{figure}
\includegraphics[width=\columnwidth]{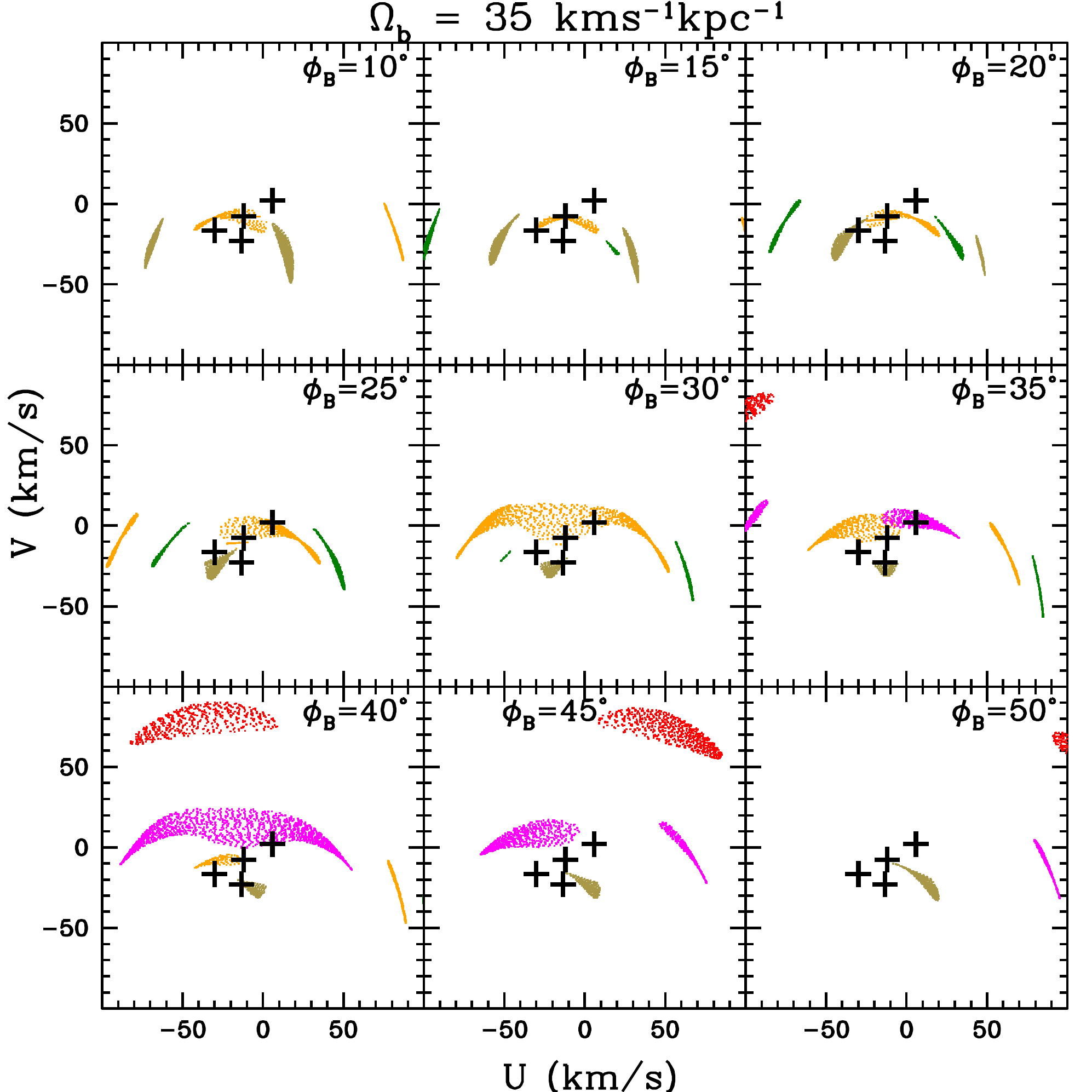}
\caption{U-V velocities with respect to the Sun of stable sections in
 resonant families with $\Omega_{\rm b}$=35 $\kmskpc$.
 The family colour is as in Fig.~\ref{figura2}. The value of the
 angle $\phi_B$ appears in each panel. The solar vicinity has a radius
 of 500 pc. The four main density maxima in Fig.~\ref{figura11}
 are shown with plus signs.}
\label{figura1ap}
\end{figure}

\begin{figure}
\includegraphics[width=\columnwidth]{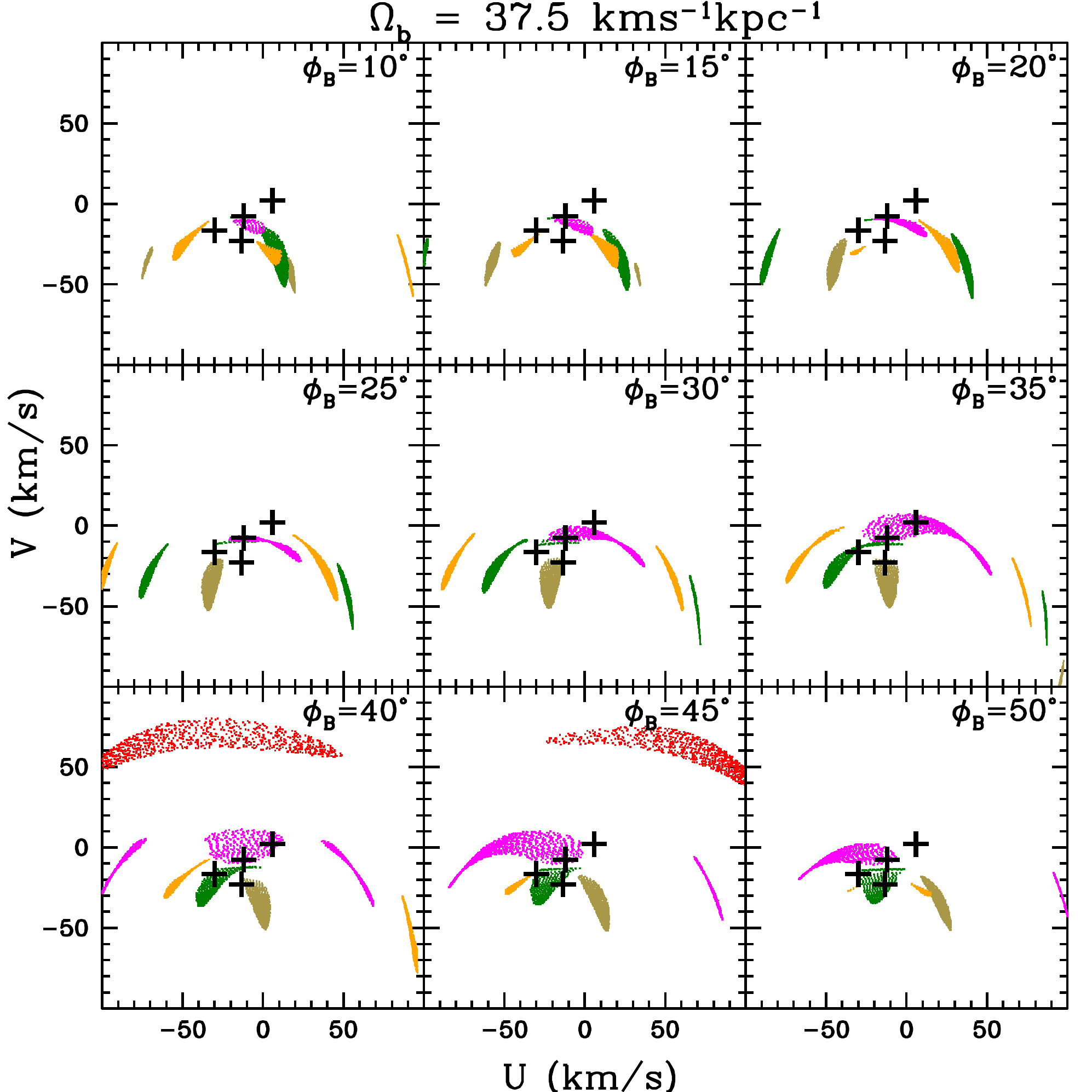}
\caption{Same as Fig.~\ref{figura1ap},
with $\Omega_{\rm b}$=37.5 $\kmskpc$ and different values of the
angle $\phi_B$.}
\label{figura2ap}
\end{figure}

\begin{figure}
\includegraphics[width=\columnwidth]{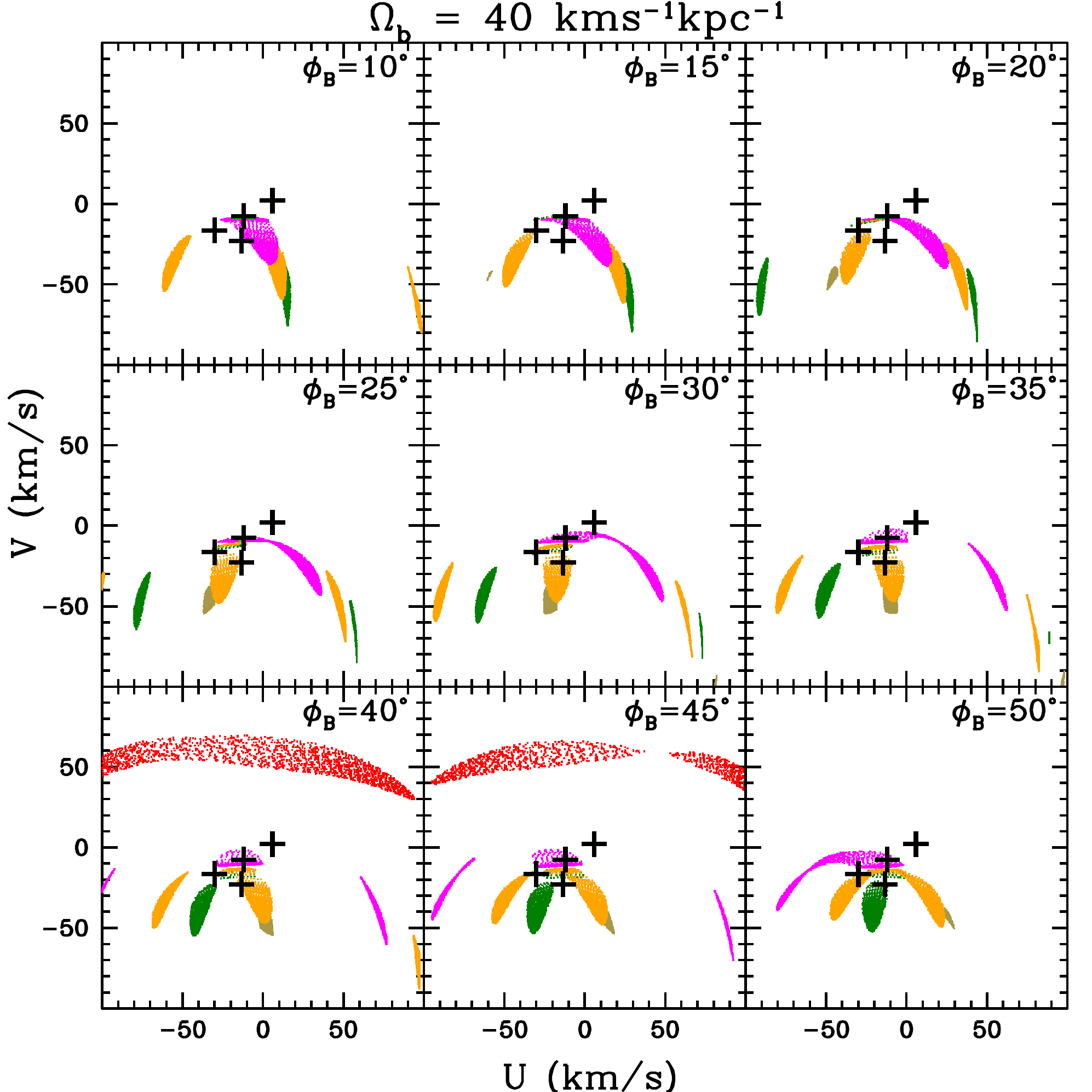}
\caption{Same as Fig.~\ref{figura1ap},
with $\Omega_{\rm b}$=40 $\kmskpc$ and different values of the
angle $\phi_B$.}
\label{figura3ap}
\end{figure}

\begin{figure}
\includegraphics[width=\columnwidth]{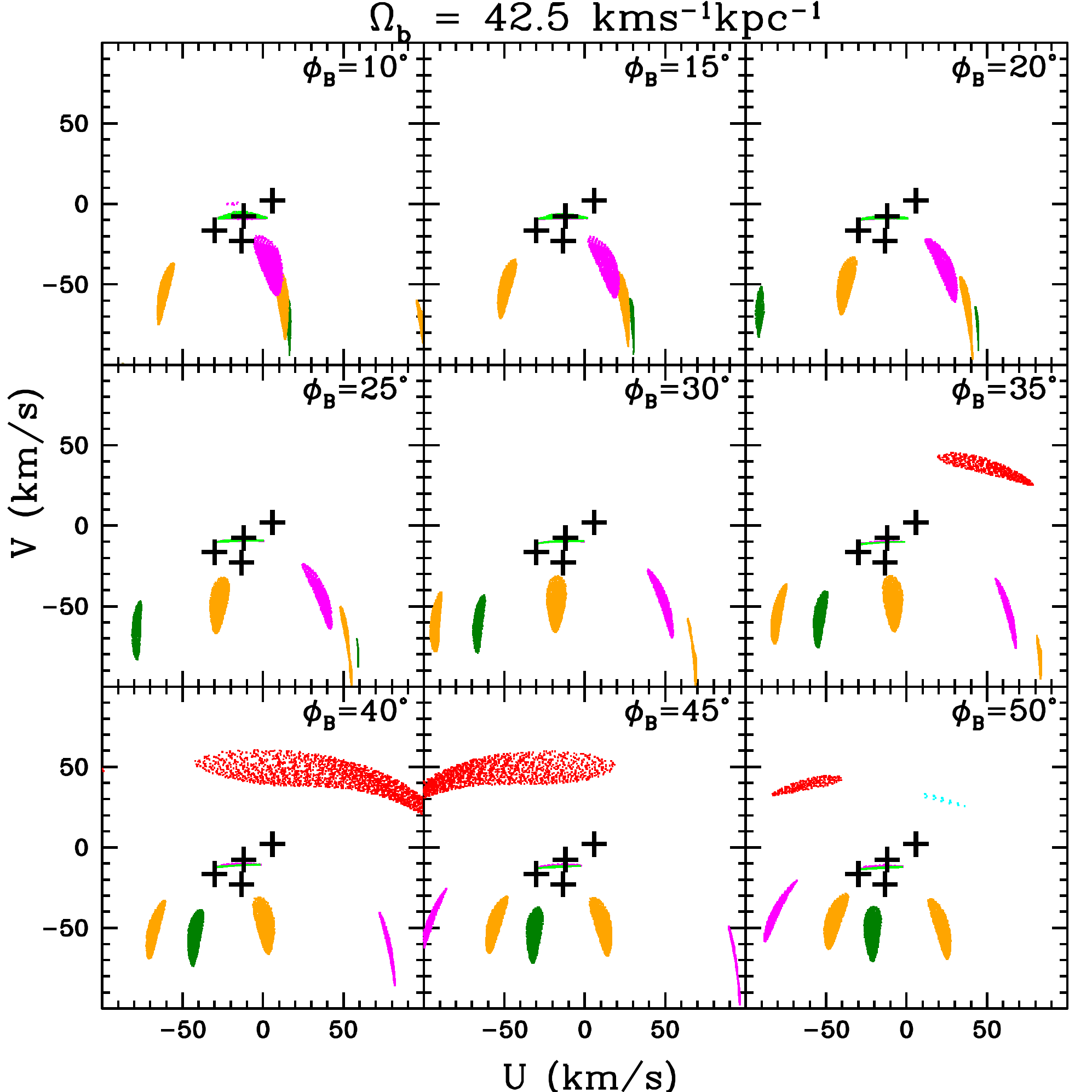}
\caption{Same as Fig.~\ref{figura1ap},
with $\Omega_{\rm b}$=42.5 $\kmskpc$ and different values of the
angle $\phi_B$.}
\label{figura4ap}
\end{figure}

\begin{figure}
\includegraphics[width=\columnwidth]{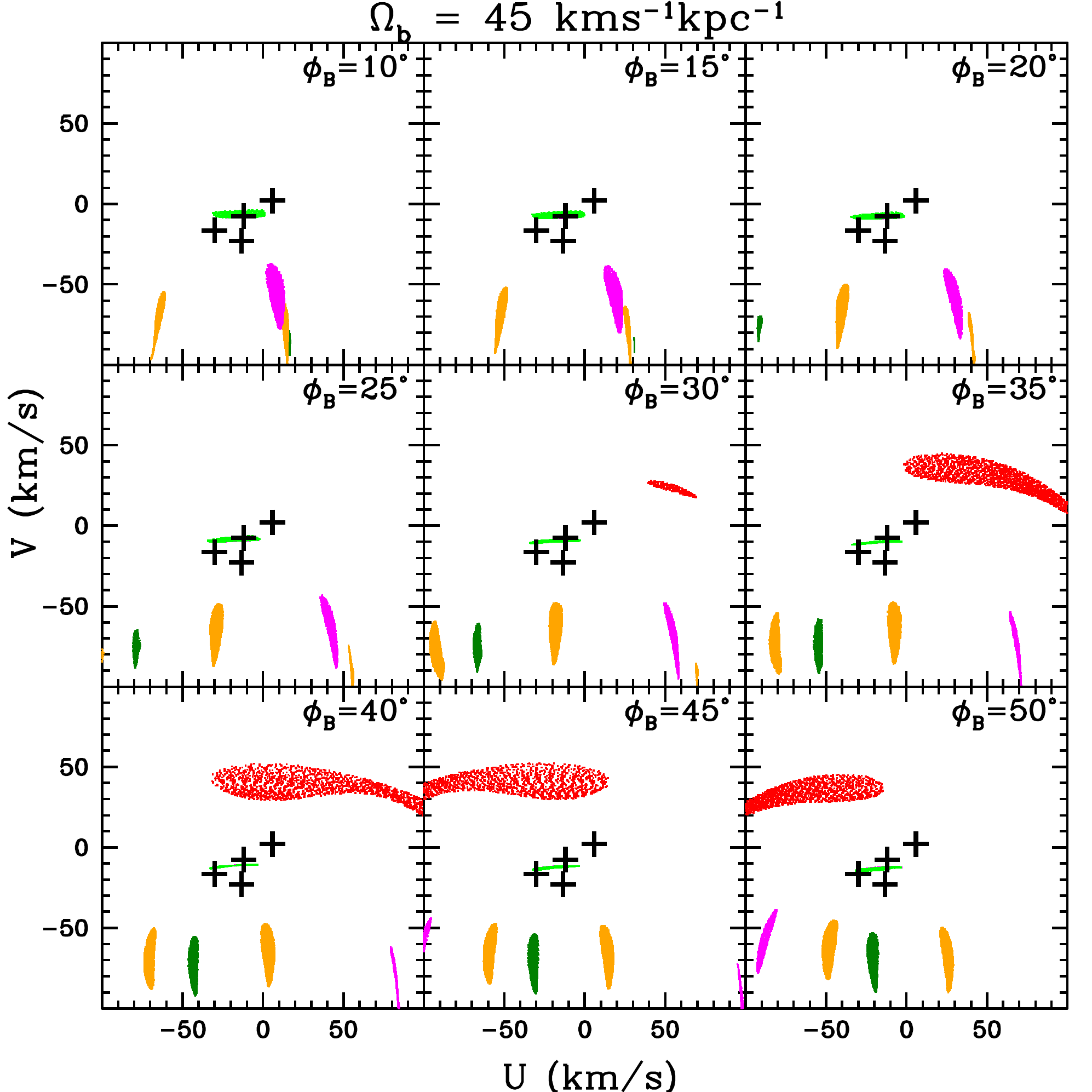}
\caption{Same as Fig.~\ref{figura1ap},
with $\Omega_{\rm b}$=45 $\kmskpc$ and different values of the
angle $\phi_B$.}
\label{figura5ap}
\end{figure}

\begin{figure}
\includegraphics[width=\columnwidth]{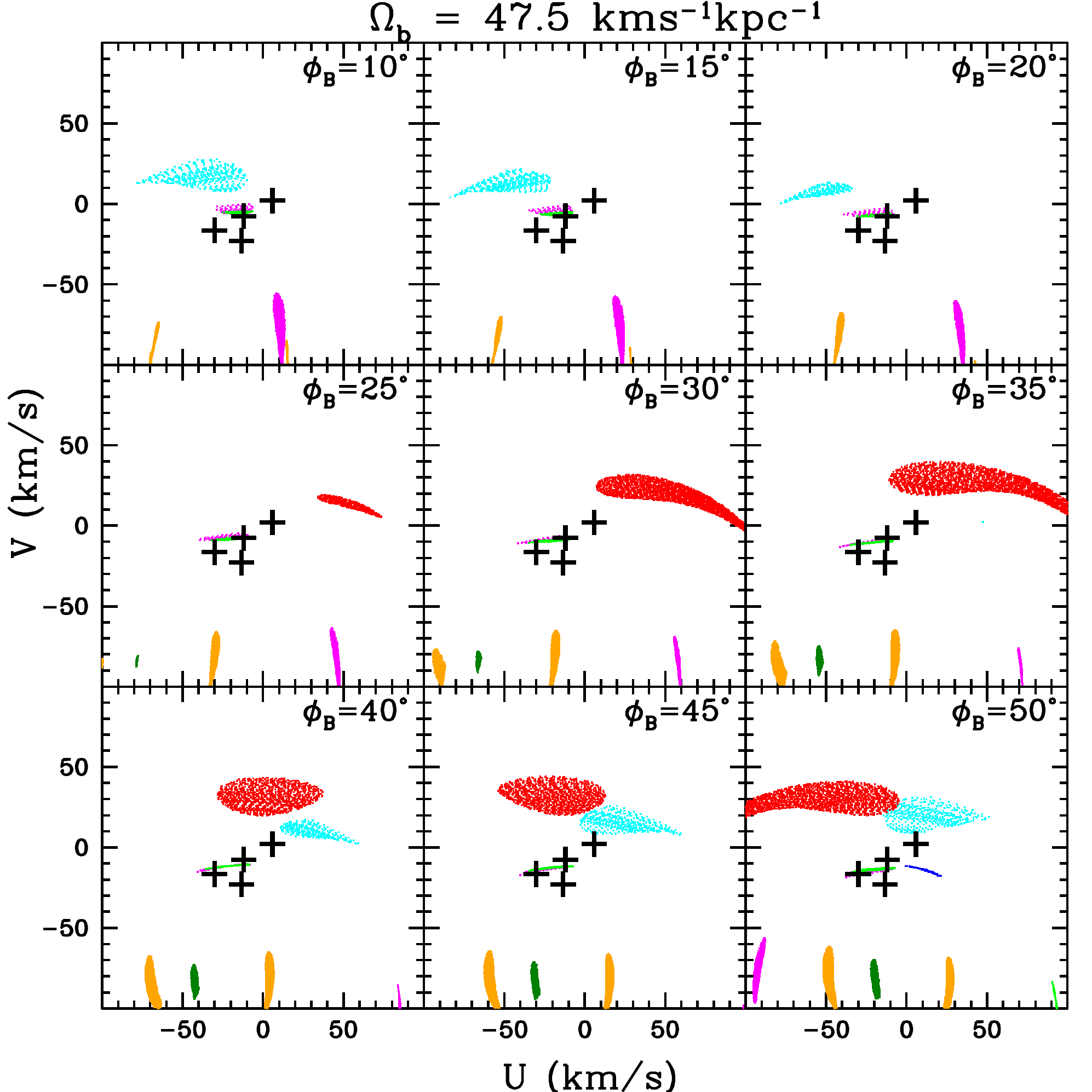}
\caption{Same as Fig.~\ref{figura1ap},
with $\Omega_{\rm b}$=47.5 $\kmskpc$ and different values of the
angle $\phi_B$.}
\label{figura6ap}
\end{figure}

\begin{figure}
\includegraphics[width=\columnwidth]{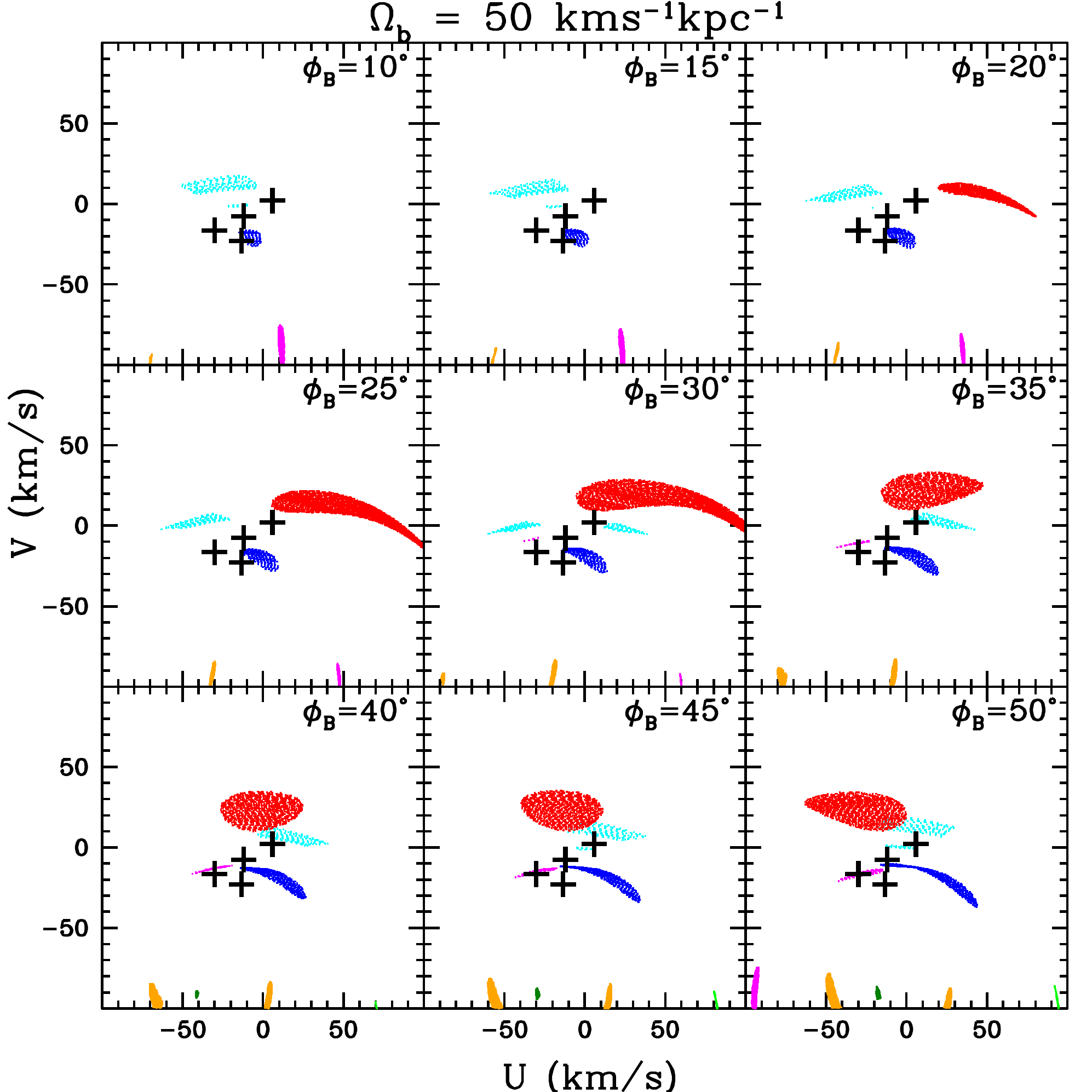}
\caption{Same as Fig.~\ref{figura1ap},
with $\Omega_{\rm b}$=50 $\kmskpc$ and different values of the
angle $\phi_B$.}
\label{figura7ap}
\end{figure}

\begin{figure}
\includegraphics[width=\columnwidth]{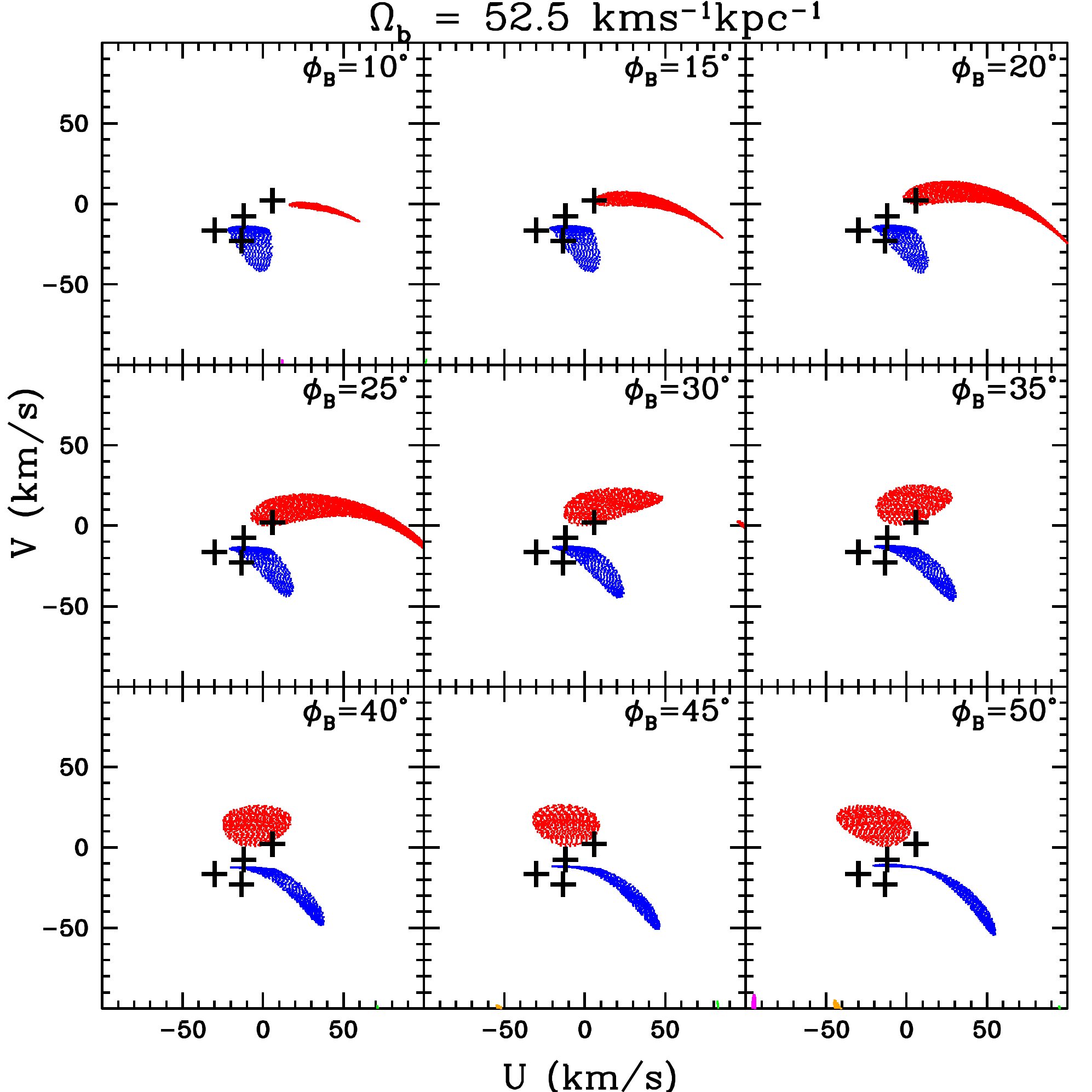}
\caption{Same as Fig.~\ref{figura1ap},
with $\Omega_{\rm b}$=52.5 $\kmskpc$ and different values of the
angle $\phi_B$.}
\label{figura8ap}
\end{figure}

\begin{figure}
\includegraphics[width=\columnwidth]{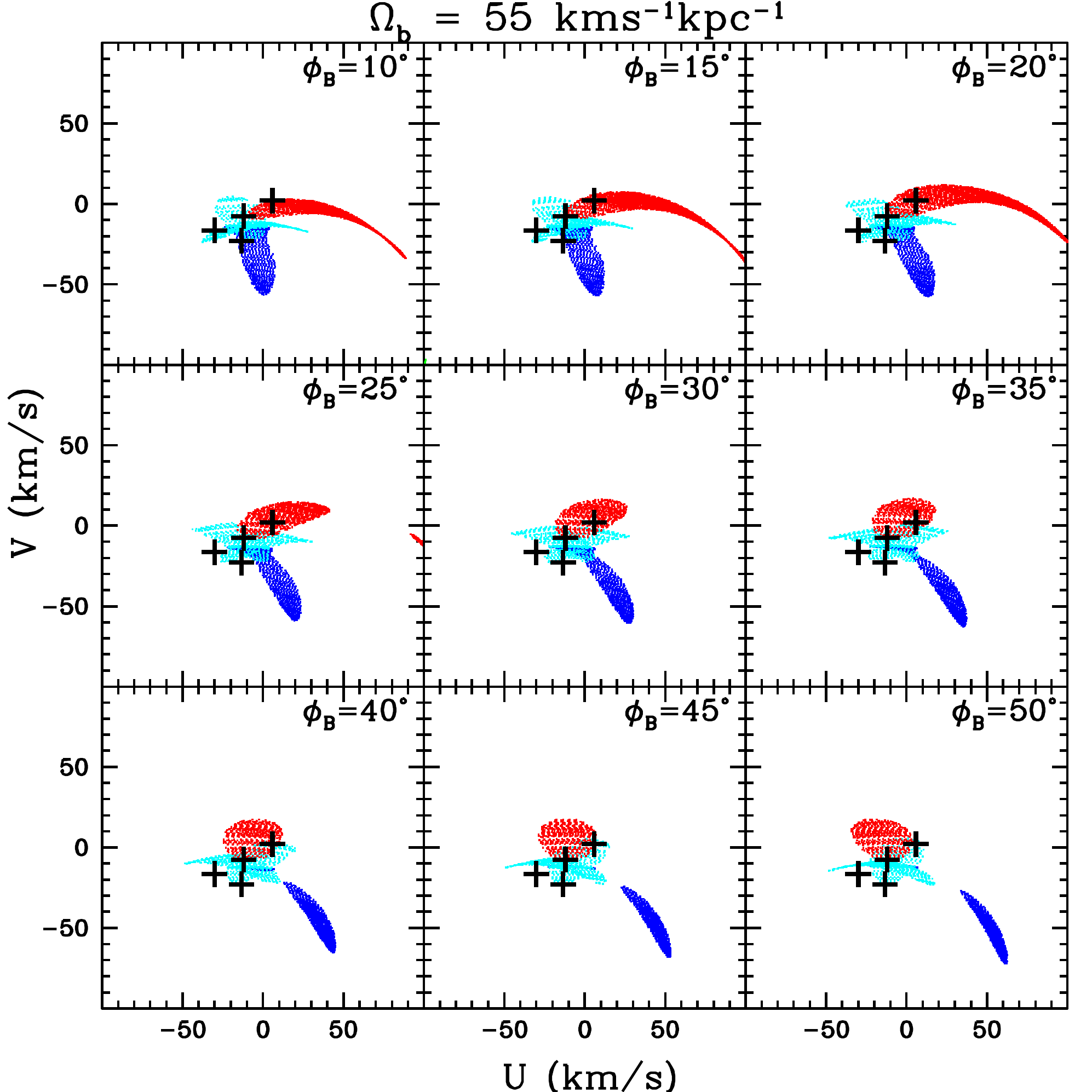}
\caption{Same as Fig.~\ref{figura1ap},
with $\Omega_{\rm b}$=55 $\kmskpc$ and different values of the
angle $\phi_B$.}
\label{figura9ap}
\end{figure}

\begin{figure}
\includegraphics[width=\columnwidth]{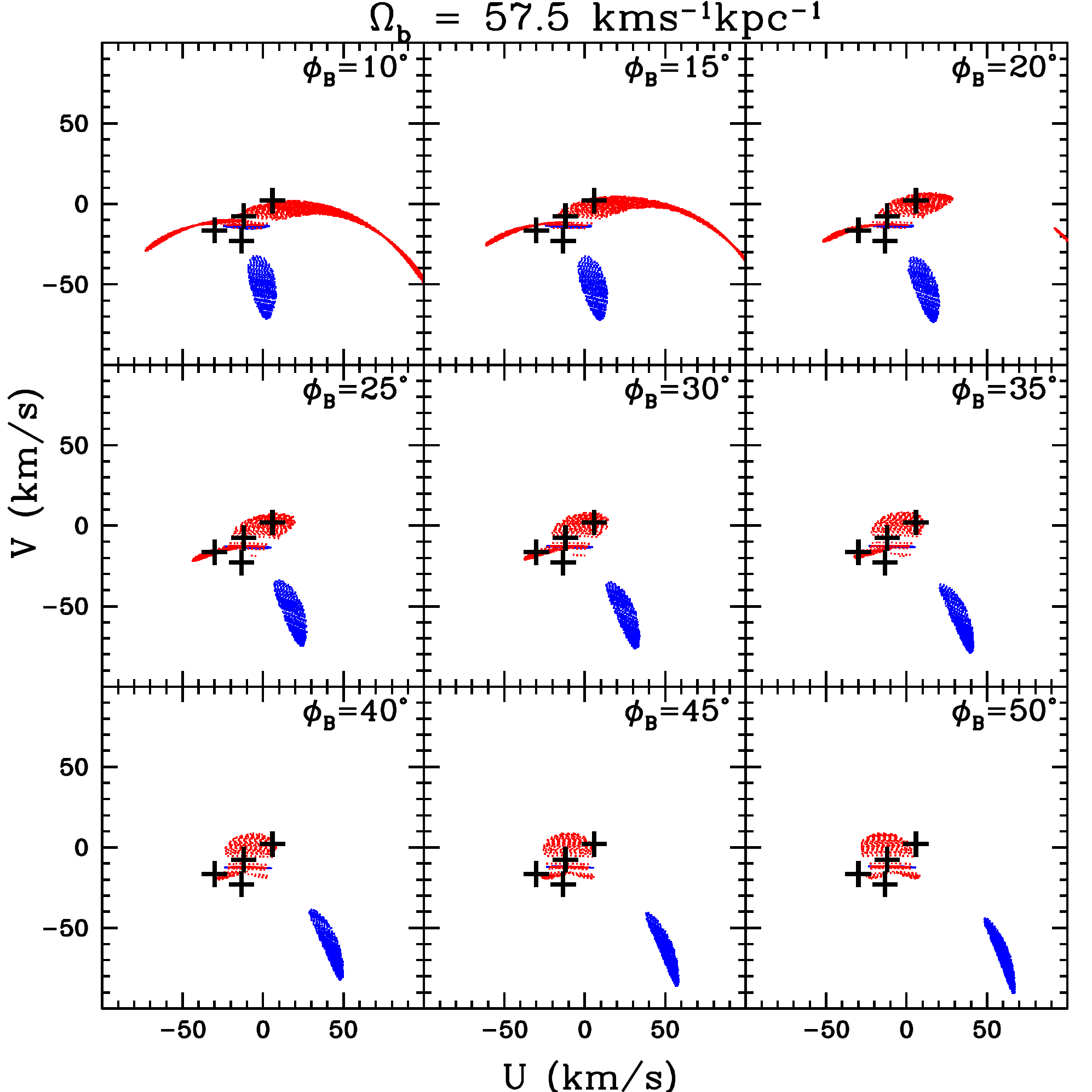}
\caption{Same as Fig.~\ref{figura1ap},
with $\Omega_{\rm b}$=57.5 $\kmskpc$ and different values of the
angle $\phi_B$.}
\label{figura10ap}
\end{figure}




\bsp	
\label{lastpage}
\end{document}